\documentclass[12pt]{article}
\usepackage{latexsym,epsfig,graphicx,epstopdf,amsmath,amssymb,amscd,multirow,chicago,psfrag,paralist,dsfont,url}
\usepackage[titletoc]{appendix}

\usepackage[american]{babel}
\usepackage{varwidth}
\usepackage{verbatim}
\usepackage{bbm}
\usepackage{tikz}
\usetikzlibrary{shapes.geometric, arrows}

\usepackage{ctable}
\newcommand{\thetavec}{{\boldsymbol{\theta}}}
\newcommand{\veps}{\varepsilon}

\newcommand{\betavec}{{\boldsymbol{\beta}}}

\newcommand{\E}{\mathds{E}}

\newcommand{\xvec}{\boldsymbol{x}}
\newcommand{\svec}{\boldsymbol{s}}
\newcommand{\tvec}{\boldsymbol{t}}

\newcommand{\boldD}{\mathbf{\mathcal{D}}}
\newcommand{\X}{\mathcal{X}}
\newcommand{\Hs}{\mathcal{H}}

\newcommand{\err}{\text{err}}

\begin{document}

\title{\vspace{-3ex} Bridging the Data Gap in AI Reliability Research and Establishing DR-AIR, a Comprehensive Data Repository for AI Reliability}

\author{
Simin Zheng$^{1}$, Jared M. Clark$^{1}$, Fatemeh Salboukh$^{2}$, Priscila Silva$^{3}$, \\ Karen da~Mata$^{3}$, Fenglian Pan$^{4}$,
Jie Min$^{5}$, Jiayi Lian$^{6}$, Caleb B. King$^{7}$,\\ Lance Fiondella$^{3}$, Jian Liu$^{4}$, Xinwei Deng$^{1}$, and Yili Hong$^{1}$\\[1.5ex]
{\small $^{1}$Department of Statistics, Virginia Tech, Blacksburg, VA 24061}\\
{\small $^{2}$Dept. of Engineering \& Applied Science, Univ. of Massachusetts, Dartmouth, MA 02747}\\
{\small $^{3}$Dept. of Electrical \& Computer Engineering, Univ. of Massachusetts, Dartmouth, MA 02747}\\
{\small $^4$Department of Systems \& Industrial Engineering, University of Arizona, Tucson, AZ 85721}\\ 	
{\small $^5$Department of Mathematics \& Statistics, University of South Florida, Tampa, FL 33620}\\
{\small $^{6}$Quantitative Analytics, Wells Fargo, Charlotte, NC 28262}\\
{\small $^{7}$JMP Division, SAS, Cary, NC 27513}
}


\date{}

\maketitle

\begin{abstract}
Artificial intelligence (AI) technology and systems have been advancing rapidly. However, ensuring the reliability of these systems is crucial for fostering public confidence in their use. This necessitates the modeling and analysis of reliability data specific to AI systems. A major challenge in AI reliability research, particularly for those in academia, is the lack of readily available AI reliability data. To address this gap, this paper focuses on conducting a comprehensive review of available AI reliability data and establishing DR-AIR: a data repository for AI reliability. Specifically, we introduce key measurements and data types for assessing AI reliability, along with the methodologies used to collect these data. We also provide a detailed description of the currently available datasets with illustrative examples. Furthermore, we outline the setup of the DR-AIR repository and demonstrate its practical applications. This repository provides easy access to datasets specifically curated for AI reliability research. We believe these efforts will significantly benefit the AI research community by facilitating access to valuable reliability data and promoting collaboration across various academic domains within AI. We conclude our paper with a call to action, encouraging the research community to contribute and share AI reliability data to further advance this critical field of study.


\end{abstract}

\newpage

\section{Introduction}\label{sec:introudction}
\subsection{Motivation and Objectives}

Artificial intelligence (AI) technology and systems have been advancing at an unprecedented pace. Examples of AI systems include autonomous systems such as self-driving cars, drones, and industrial robots (e.g., \shortciteNP{soori2023artificial}); natural language processing (NLP) systems like advanced conversational AI systems (e.g., ChatGPT), virtual assistants (e.g., Alexa), and language translation tools \shortcite{mohamed2024impact}; computer vision systems like facial recognition \shortcite{nawaz2020artificial}; and AI systems in healthcare \shortcite{koski2021ai}. While there is considerable excitement surrounding AI technology, ensuring the reliability of these systems is essential for building public confidence in their wide adoption. Reliability issues can result in significant losses and even catastrophic failures, highlighting the importance of AI reliability.

Ensuring the reliability of AI systems requires the modeling and analysis of reliability data specific to these emerging technologies. However, a significant challenge in AI reliability research is the lack of readily available data, which arises from several factors. First, as AI technology is still in its rapidly evolving early stages, development efforts often prioritize performance metrics like accuracy and speed over reliability and other important metrics, such as robustness. Second, while industries may generate and utilize reliability data through applied testing, the academic sector often lacks comparable testbeds, resulting in a data gap. Third, data sharing presents additional challenges. Industrial reliability data are often proprietary and sensitive, limiting access for academic researchers. These factors collectively create challenges for the AI reliability research community, particularly in academia, where access to real data is important for effectively modeling and analyzing AI reliability.

Therefore, this paper seeks to address the data gap in AI reliability research by introducing key concepts related to AI reliability data, providing a comprehensive review of the currently available data, and establishing a novel public data repository to facilitate data sharing. This work is essential because data is a cornerstone of research. Researchers in AI reliability come from diverse fields, including machine learning (ML), statistics, electrical engineering, computer engineering, industrial systems engineering, and other related disciplines. While datasets exist, they are fragmented across disciplines, and inconsistent terminology often complicates their integration. In addition, some public data repositories, such as \citeN{UCIRepo} and \citeN{Kaggle}, provide access to datasets suitable for statistical analysis and ML model building. However, they do not specifically focus on AI reliability analysis, which is the primary focus of our study.

Before collecting data, it is important to identify the appropriate metrics for evaluating AI reliability. In this paper, we discuss key measurements and data types for assessing AI reliability, along with methodologies for data collection, such as the design of experiments (DoE) and accelerated life tests (ALT). We also provide a detailed introduction to currently available datasets, complemented by illustrative examples. Finally, we establish DR-AIR, a data repository for AI reliability. DR-AIR provides access to a diverse collection of curated datasets designed to support and advance research in AI reliability. In addition, we outline the structure of the DR-AIR repository and demonstrate its practical applications. We believe these efforts will greatly benefit the AI research community by improving access to valuable data for reliability analysis.

\subsection{Literature Review and The Contribution}

AI systems have become increasingly popular and widely used across many fields. With advancements in AI technology, demonstrating the reliability of these systems is essential for their confident use. \shortciteN{werner2022leveraging} proposed a framework for developing, qualifying, and releasing reliable and assured AI systems by applying design for reliability tools and techniques during the design and development phases. \shortciteN{blood2023reliability} highlighted that traditional reliability tools need to be transformed to address the reliability of AI systems. \shortciteN{hong2023statistical} provided a compressive discussion on statistical reliability for AI systems. Existing research has made significant contributions in the field of AI systems, highlighting the importance of exploring the data used in AI reliability research.

To further emphasize the importance of data exploration for AI reliability research, several data collection methods from the field of traditional reliability analysis have already been investigated. Some of these methods could be further extended to AI system reliability studies. \shortciteN{smith2021reliability} illustrated a method for failure data collection, as well as a structured approach to recording the data using a formal document from the field, which can be used for reliability analysis. \shortciteN{meeker2022statistical} introduced data collection strategies that can be applied to planning reliability studies, as well as to data analysis and modeling in reliability research. \shortciteN{inel2023collect} developed a responsible AI methodology designed to guide data collection, which can be used to assess the robustness of data used for AI applications in the real world. However, a gap remains in the detailed introduction of data collection methods specifically for AI system reliability research.

Although AI system reliability research has emerged as a growing field in recent years, the availability of data for this research remains limited. For AI system reliability analysis, \shortciteN{hong2023statistical} used the public \citeANP{AIIncidentDB} database~(2021), which primarily collects AI incidents from news reports, and applied a text mining method to identify variables that contribute the most to AI incident data to illustrate the importance of AI reliability. \shortciteN{MinHongKingMeeker2020} and \shortciteN{Zheng2023-testplan} both used publicly accessible data from the California Department of Motor Vehicles (DMV). Specifically, \shortciteN{MinHongKingMeeker2020} focused on parametric and non-parametric models to describe disengagement events from autonomous vehicles, while \shortciteN{Zheng2023-testplan} focused on test planning for reliability assurance tests. \shortciteN{Panetal2024} used data from a physics-based AV simulation platform to demonstrate the reliability prediction performance and interpretability of an error propagation model. In terms of assessing the robustness of advanced ML algorithms, \shortciteN{Lianetal2021Robustness} collected prediction performance results by conducting a comprehensive set of mixture experiments to assess the robustness classification algorithms. \shortciteN{Faddietal2024} conducted experiments to collect datasets that capture the behavior of machine learning image classifiers on both clean and perturbed inputs to evaluate the reliability of AI algorithms. Some available data can be used for AI reliability studies; however, it remains limited and requires further exploration.

In summary, the currently available datasets for AI reliability research remain limited. Therefore, we aim to address this gap by creating a publicly accessible repository focused on AI reliability. The established public online repository provides several contributions. First, it is a valuable resource for AI reliability researchers by providing public access to reliability data, which can serve as a starting point for AI reliability research. Second, it facilitates communication between ML and reliability researchers, as well as researchers from other fields, by enabling collaboration across various academic domains. Third, highlighting the role of AI reliability in ensuring safety can help attract a broader range of researchers from the academic community, fostering further research and the development of new methods in this emerging field.

\subsection{Overview}

The rest of the paper is organized as follows. Section~\ref{sec:measurement.data.type} discusses commonly used metrics and measurements for AI reliability, along with their associated data types. Section~\ref{sec:design.data.collection} discusses methods and strategies for effective data collection to support AI reliability. Section~\ref{sec:datasets.illustration} introduces the datasets we have collected, providing detailed examples of their use in modeling and analysis. Section~\ref{sec:DR.AIR.setup} explains the setup of the online repository, including its data format and usage guidelines. Finally, Section~\ref{sec:concluding.remarks} offers concluding remarks and highlights calls to action for advancing AI reliability data collection, modeling, and analysis.

\section{Measurements and Data Types}\label{sec:measurement.data.type}
In this section, we provide a comprehensive discussion on measurement and data types for AI reliability, including covariate information, which plays an important role in addressing the data gap in AI reliability research.

\subsection{AI Reliability Measurement}

We first focus on addressing how AI reliability can be measured. The work of \shortciteN{hong2023statistical} outlines the process of defining AI reliability metrics. Here, failure rate, event rate, and error rate are all noted as possibilities for measuring the reliability of an AI system. AI systems are built with an intended purpose. In general, an AI system is reliable if it can perform its purpose for a ``long'' period of time, which corresponds to the formal definition of reliability. Reliability is defined as the ability of a system (or component) to consistently perform its intended function without failure over a specified period under specified conditions.

Some AI systems may experience failures over time. For AI reliability, we need to be aware of both hardware and software failures. Hardware failures occur when the physical components of a system no longer work. A malfunctioning GPU would be an example of a hardware failure. On the other hand, a software failure occurs when the AI system fails to fulfill its intended purpose successfully. Note that not all failures prevent the system from being used in the future. In this context, we use ``failure'' as a broad term, allowing for the possibility of multiple failures in the same unit. When failures are clearly defined, possible reliability metrics may include whether a system failed, how long it operated before failure, or, in the case of multiple failures, the rate of failure events. For example, autonomous vehicles (AVs) are one area where AI reliability is currently under investigation. The work of \shortciteN{MinHongKingMeeker2020} considers disengagement as its ``failure'' mode. A disengagement event occurs when the AV exits autonomous mode and gives control to the driver. \shortciteN{Panetal2024} defines the failure of AVs in terms of errors in the perception system. Both studies have different but clearly defined metrics for failure. These studies can each provide different but meaningful insights into the reliability of AI systems employed in AVs.

Different from traditional reliability, failure in many AI systems is not easily defined in terms of time. Generative AI systems are often assessed based on the accuracy of their output. For example, chatbots are often able to provide code for programming problems. The AI system can then be evaluated based on how accurately it executes the task specified in the prompt. For these systems, a time component may be less meaningful. Instead, we focus on assessing the overall error rate in order to determine how often the model is successfully performing the required task. The work of \shortciteN{Lianetal2021Robustness} considers the accuracy of classification from AI algorithms. In this case, classification accuracy metrics are used to measure the reliability of the AI systems under investigation. For another example, adversarial networks play a large role in current machine learning research. \shortciteN{Faddietal2024} also considers AI classification algorithms. In this case, the authors employ adversarial attacks which aim to cause misclassification from the AI system. AI reliability is measured in multiple ways. First, the times of successful adversarial attacks were recorded. Additionally, total failure counts were measured for each run of the experiment.

\subsection{Data Types}
Next, we discuss the data types used in reliability measurement. The data type of the response variable is particularly important, as it determines the appropriate statistical models for subsequent analysis. Various data types are used in reliability studies, including binary data, count data, continuous measurement data, time-to-event data, recurrent event data, and degradation data. In the following sections, we provide a detailed discussion of each.

Binary responses are commonly used in early reliability studies, where outcomes are recorded as pass/fail. These data are modeled using a Bernoulli distribution. In the absence of covariates, the distribution can be parameterized with a shared probability of event occurrence. When covariates are present, the generalized linear model (GLM) is typically used for analysis (e.g., \citeNP{McCullaghNelder1999}). The most commonly applied GLMs for binary data are logistic and probit regression. In AI reliability studies, binary outcomes can also be relevant.

Count data typically arises when the response represents the number of events (e.g., failures) occurring within specific time and unit constraints. These data are typically modeled as following a Poisson distribution. In the presence of covariates, these data can also be analyzed with the use of GLMs. The failure count data of \shortciteN{Faddietal2024} is an example of count data. In this example, the AI system classifies images, and we record the number of misclassified images for the run of the adversarial network experiment.

Continuous measurement responses are less common in traditional reliability studies but are more prevalent in AI reliability. These responses take values on the real line and are often modeled using a normal distribution with unknown location and scale parameters. However, a normal assumption may not always be appropriate, requiring functional transformations to ensure unbounded support or reduce skew. In cases of bounded responses, normal fits may still be reasonable if probabilities beyond the bounds are negligible. An example is provided in \shortciteN{Lianetal2021Robustness}, where AI classification reliability is analyzed using the mean area under the curve (AUC) and the log standard deviation of AUC, both of which are continuous measures of classification accuracy.

Time-to-event (or time-to-failure) data is crucial in reliability analysis, recording the time until an event for each unit (\shortciteNP{meeker2022statistical}). Some units may not experience an event during observation, resulting in right-censored data. Although censored observations do not provide exact failure times, they still contribute to likelihood estimation, impact assessment, and inference. Time-to-event data is fundamental in traditional reliability studies, with common models assuming log-normal or Weibull distributions, or more generally, the log-location-scale families. For example, \shortciteN{Faddietal2024} includes time measurements of adversarial attack successes. Compared to binary and count data, failure time data is more informative, allowing the reconstruction of failure indicators.

Recurrent event data, like time-to-event data, focuses on the distribution of event times but differs in that units can experience multiple events. This distinction requires different models, often using point processes like the non-homogeneous Poisson process (NHPP) to model the time between events. \shortciteN{MinHongKingMeeker2020} provides an example, with data originally collected from the  \citeN{CAdriving} and cleaned for analysis, recording disengagement events in AVs -- an instance of system-level test data, where failure occurs when the system fails its intended function. Similarly, \shortciteN{Panetal2024} examines AV failures but defines them as AI detection errors, representing module-level test data, where failure occurs when an AI component misperforms its task.

Finally, we consider degradation data (\shortciteNP{meeker2022statistical}), which is common in traditional reliability but not yet seen in AI reliability. Unlike hard failures that render a unit inoperable, degradation occurs gradually. Binary failure data can be generated by defining a ``soft'' failure when degradation surpasses a threshold, but this approach loses information, as the full degradation path cannot be reconstructed without strong assumptions like linearity. Examples include tire tread wear in vehicles and efficiency loss in physical systems.

\subsection{Covariates}

In traditional reliability analysis, covariates are useful because they help explain more variability in the responses and enhance the predictability of future outcomes. Typically, traditional reliability data, such as those from ALT, do not include long lists of variables. The accelerating variables are usually limited to one or two, such as temperature or voltage. However, in AI reliability, a wide variety of covariates can be collected for analysis, offering more opportunities for statistical modeling and analysis. For example, in AI/ML models, the type of algorithm becomes a factor in the dataset, and this information is included as a covariate. This is illustrated in \shortciteN{Lianetal2021Robustness}, where different algorithms are compared in terms of robustness to unbalanced data. Similarly, the operating company of an AI system plays a comparable role in \shortciteN{MinHongKingMeeker2020}, as different companies may use different systems, and we seek to understand how their vehicles compare. \shortciteN{MinHongKingMeeker2020} also include mileage information as a covariate. When investigating algorithms through simulation, simulation settings can also serve as covariates. For instance, \shortciteN{Faddietal2024} include the percentage of adversarial attacks created by two different algorithms in the dataset. Likewise, \shortciteN{Lianetal2021Robustness} include the proportion of each class used in the training dataset as the covariate.

There are several general model strategies to incorporate covariates to explain the response. First, for observation $i$, we denote the covariate information in a $p \times 1$ vector, $\xvec_i$. For categorical covariates, this often means a one-hot encoding of the variable, as mentioned in \shortciteN{dahouda2021deep}.
For continuous measurement data, the inclusion of covariate information is rather simple. Let $\mu_i$ be the mean for the assumed normal distribution of the data. Then,
$$\mu_i=\xvec_i'\betavec, $$
where $\betavec=(\beta_{1}, \dots ,\beta_{p})'$ is the vector of coefficient parameters, and $p$ is the number of coefficients. This defines the typical linear model, which allows for the leveraging of covariate information. Covariate information can be included in a similar manner for binary data and count data under the GLM framework. In this case the mean $\mu_i$ is linked to the linear predictor $\xvec_i'\betavec$ through a link function $g(\cdot)$, that is,
$g(\mu_i) = \xvec_i'\betavec$.

Then, a regression approach is commonly used to incorporate covariate information when analyzing time-to-event data, specifically in the accelerated failure time (AFT) model. Let $t_i$ be the time to event, the AFT model is:
$$
\log(t_i)=\xvec_i'\betavec+\sigma\veps_i,
$$
where $\sigma$ is the scale parameter of the error term $\veps_i$, which follows a standard location-scale distribution. The cumulative damage model (e.g., \citeNP{HongMeeker2013}) can be used if there are time-varying covariates.

Third, the approach to modeling recurrent event data is analogous to the method used for time-to-event data. However, we model the intensity function, $\lambda_i(t)$, as follows:
$$
\lambda_i(t) = \lambda_0(t) \exp(\xvec_i' \betavec).
$$
In this case, we assume the NHPP and treat $\lambda_i(t)$ as the intensity function, leaving $\lambda_0(t)$ as the baseline intensity function (BIF). Time-varying covariates can be incorporated similarly. Nevertheless, many novel models for analyzing AI reliability data are currently the focus of ongoing research.

\section{Designs and Methods for Data Collection}\label{sec:design.data.collection}
In this section, we provide a comprehensive description on AI reliability data collection, covering key aspects such as the two main data sources (laboratory vs. field), the two methods of data collection (virtual vs. physical), and relevant statistical techniques, including DoE and ALT.

\subsection{Laboratory Tests and Field Tracking Studies} \label{sec:lab.field}

Data collection serves as the foundation for AI reliability research. There are various designs and methods for data collection. As \shortciteN{karunarathna2024crucial} pointed out, choosing the appropriate data collection method is important and depends on the specific research questions. Traditional reliability data are collected through either laboratory tests or field tracking studies.

Data collection using laboratory tests involves gathering data under controlled experimental conditions, typically within a laboratory setting where variables and conditions are precisely regulated. Traditionally, product reliability is first tested in a laboratory environment, followed by an assessment of its reliability, leading to the generation of laboratory test data. Since most AI systems are software-based, testing them in a laboratory environment is convenient. Laboratory tests can be conducted at various levels, such as the algorithm level, module level, or system level. At the algorithm level, the test involves running the algorithm on a computer. For example, \shortciteN{Lianetal2021Robustness} and \shortciteN{Faddietal2024} evaluated the performance of CNNs in a laboratory environment. \shortciteN{pan2022quantifying} tested an AV perception system (module-level test) in a laboratory environment. \shortciteN{howard2021reliability} employed laboratory tests to collect data for evaluating the reliability and validity of a face recognition system, which can be regarded as a system-level test. Although laboratory testing can be comprehensive, its operating environment may differ from real-world scenarios. Thus, a field tracking study may be necessary.

Field studies involve collecting data outside of experimental or laboratory settings. This type of data collection is most often conducted in natural environments. The key difference with the field tracking studies method is the use of experimental methods in a ``field" situation where the data can be controlled to a limited extent, as pointed out by \shortciteN{fellows2021research}. It aims to capture more original and representative data compared to controlled laboratory tests; however, it can also be expensive and time-consuming. In the AI reliability area, the California DMV study analyzed in \shortciteN{MinHongKingMeeker2020} can be considered a field tracking study, where AVs are tested on city roads, and reliability data are collected for analysis.

\begin{figure}
	\centering
	\includegraphics[width=.85\textwidth]{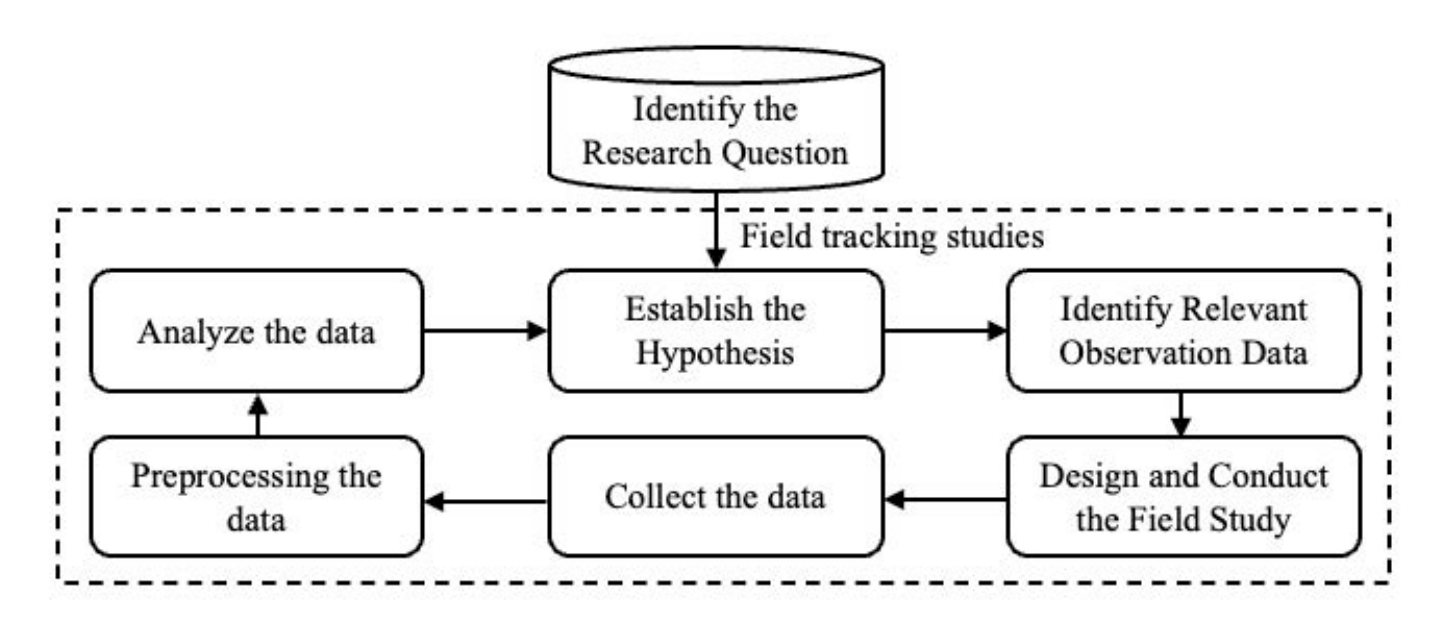}
\caption{AI reliability data collection workflow using field tracking studies.}
     \label{fig:field.study.workflow}
\end{figure}

Based on \shortciteN{gupta2022research}, we summarized and developed a typical workflow for data collection in AI reliability studies using the field tracking method. More details can be found in Figure~\ref{fig:field.study.workflow}. Specifically, before conducting field tracking studies for AI reliability research, researchers must first define the specific research question. Once clarified, they should establish a hypothesis to explain expected outcomes. Based on this hypothesis, researchers identify the relevant data to observe, guiding the design of the study. The collected data is then preprocessed based on the specific research questions. Finally, the data is processed for analysis to test the hypothesis, determining whether it should be accepted or rejected. One example of data acquisition based on a field tracking study is the predictive analysis for AVs discussed in \shortciteN{goriparthi2024ai}, where data was systematically collected from autonomous systems operating in real-world environments. The collected data included real-time telemetry (e.g., speed, temperature, vibration, and power consumption) and AI system logs.

\subsection{Virtual and Physical Tests}

Virtual and physical tests represent two forms of testing used to evaluate the reliability of AI systems. With the rapid development of technology and the digital age, virtual platforms can simulate real-world scenarios, enabling AI systems to operate under various conditions without the need for labor-intensive and time-consuming real-world data collection procedures. Virtual testing eliminates the need to set up physical environments; instead, all conditions are generated virtually using simulations or algorithms. In a virtual test, data can be collected even under simulated extreme conditions. For instance, scenarios involving AV accidents can be simulated to evaluate how the AI system operates and responds in such challenging and critical situations. In practice, various simulation platforms are available to conduct virtual tests for data collection. In recent years, Simulation of Urban Mobility has been an open-source platform for road traffic simulation, as discussed in \shortciteN{krajzewicz2010traffic}, and is widely used to evaluate traffic management AI. With advancements in innovation, more flexible sensor settings and environmental conditions have become available. Another open urban driving simulator, CARLA, introduced in \shortciteN{dosovitskiy2017carla}, provides a simulation platform that supports flexible configuration of sensor settings and environmental conditions tailored to the goals of specific research studies in autonomous driving. In addition, another open-source platform for AI systems (e.g., self-driving vehicles) is Autoware, as highlighted in \shortciteN{kato2018autoware}. Specifically, Autoware is an open-source software project designed to enable AVs with embedded systems and a user-adjustable set of self-driving modules. One application of using virtual testing for data collection to evaluate sensor-fusion-based perception systems is presented in \shortciteN{Panetal2024}, where error propagation data is generated using a physics-based simulation platform.

Despite all the advantages of virtual tests, physical tests are still necessary to validate or calibrate virtual test results. In physical tests, AI systems operate in uncontrolled real-world environments under human supervision to assess and evaluate their performance, as discussed in \shortciteN{wang2023scientific}. This approach allows AI systems to function naturally while collecting data on their performance under actual conditions. Although real-world setups for physical tests can be expensive and may pose risks to people and property, they remain important for validating AI systems in practical applications. Physical tests are essential to ensuring the robustness and safety of AI systems when deployed under real-world conditions. Typical forms of physical tests for data acquisition in AI systems are varied. First, for AVs, publicly available data from on-road testing can be used to evaluate the reliability and safety of these systems. In California, AV manufacturers are allowed to test their vehicles on public roads to observe how the AVs handle unexpected situations (e.g., disengagements) and are also required to report real-time disengagement events and collision incidents for public assessment and evaluation. This process provides a way for obtaining AV data through physical testing, as demonstrated by \shortciteN{wang2020safety}, \shortciteN{MinHongKingMeeker2020}, and \shortciteN{Zheng2023-testplan}. In addition, to evaluate the reliability of unmanned aerial vehicles (UAVs), the UAV123 Dataset, a publicly available resource, contains 123 video sequences captured through aerial photography by drones for UAV-based object tracking (\shortciteNP{taufique2020benchmarking}). One application of using physical testing to obtain data sources for UAV reliability evaluation is discussed in \shortciteN{liu2022reliable}.

\subsection{The Use of DoE and ALT}

DoE and ALT can be two useful techniques for the collection of AI reliability, which are not widely used in AI literature. DoE refers to a statistical methodology for planning, designing, and analyzing experiments (\shortciteNP{antony2023design}). In a designed experiment, intentional changes are applied to input variable(s) to observe the corresponding effects on the output(s). DoE serves as a powerful approach for data collection, enabling researchers to identify treatments that produce specific outcomes (e.g., establishing cause-and-effect relationships), as described in \shortciteN{thomas2022research}.

DoE can be used in various ways for data collection. First, in traditional statistical reliability analysis, DoE can be a structured approach for planning and designing experiments tailored for data collection, as highlighted by \shortciteN{anderson2023designed}. Since the relationship between factors and the responses are complicated in AI reliability, the idea of space-filling can be useful to explore the input region. Space-filling designs such as minimax distance designs, maximin distance designs, and Latin hypercube designs, are summarized by \shortciteN{joseph2016space}. For example, consider the maximin Latin hypercube design (MmLHD) proposed by \shortciteN{morris1995exploratory}. Let $\X$ represent the experimental input region, and let $p$ denote the number of factors involved in the experimental design. Note that the experimental region is scaled to a unit hypercube, defined as $\X = [0,1]^{p}$. Let $\boldD = \{\xvec_{1}, \ldots, \xvec_{n}\}$ as the experimental design, where each designed data input $\xvec_{i} \in [0,1]^{p}$. Based on \shortciteN{morris1995exploratory}, the following criterion can be used to search for MmLHDs, which are applicable for designed data collection:
\begin{align}\label{eqn:MmLHD}
\min_{\boldD} \left\{ \left( \sum_{i=1}^{n-1} \sum_{j=i+1}^{n} \frac{1}{d^k(\xvec_i, \xvec_j)} \right)^{1/k} \right\},
\end{align}
where $d(\svec, \tvec) = \left( \sum_{i=1}^n \lvert s_i - t_i \rvert^m \right)^{1/m}$.

Next, in the evaluation of algorithm robustness in ML, DoE can also serve as a method for data acquisition, as emphasized by \shortciteN{freeman2023design}. As highlighted by \shortciteN{cody2022systematic}, existing datasets can be split into training and testing sets by leveraging combinatorial coverage. This approach can be used to provide data inputs for testing the generalizability of AI algorithms. One typical application of using DoE for data collection is presented in \shortciteN{Lianetal2021Robustness}, which considers a modified simplex centroid design for mixture experiments to test AI algorithms in predicting performance.

ALT is another method for data acquisition in a timely manner for AI reliability analysis. For some applications in AI systems, it could take months or years to collect enough data for reliability assessment under the normal use condition. For such applications, it is essential to use ALT to gather data in an accelerated way. A comprehensive introduction to traditional ALT is available at \shortciteN{escobar2006review}. To convey the main idea of ALT modeling and analysis, we introduce the parametric accelerated model commonly used in reliability modeling for ALT. Let $t_{0}$ represent the failure time under normal operating conditions and $t_{s}$ represent the failure time under stress conditions. The relationship between the two failure time scales, involving the acceleration factor $A_{F}$, is given by:
\begin{align}\label{eqn:relationship}
t_{0}=A_{F}t_{s},
\end{align}
where $$A_{F} = \frac{L_{N}}{L_{A}},$$
with $L_{N}$ representing AI system life under normal conditions and $L_{A}$ representing AI system life under accelerated stress conditions.

The cumulative distribution function (CDF) has the following relationship under two different conditions:
\begin{align}\label{eqn:cdf}
F_{0}(t)=F_{s}\left(\frac{t}{A_{F}}\right),
\end{align}
where $F_{0}(\cdot)$ is the CDF under normal conditions, and $F_{s}(\cdot)$ is the CDF under accelerated stress conditions.

As described by \shortciteN{meeker2022statistical}, data from tests conducted at high levels of accelerating variables (e.g., use rate, aging rate, or stress levels) are extrapolated through a physically motivated model. This process provides estimates of the system's lifespan under lower levels of the accelerating variables. One widely used form is the Arrhenius model, when temperature is the acceleration variable. The effect of temperature on the product is often modeled using the Arrhenius model:
\begin{align}\label{eqn:Arrhenius}
r=A\exp\left({-\frac{E_{a}}{kT}}\right),
\end{align}
where $r$ is the reaction rate, $A$ and $E_{a}$ are unknown constant. Also, $k$ is the Boltzmann constant, and $T$ is temperature in Kelvin.

Related to ALT for AI systems, \shortciteN{hong2023statistical} discussed various acceleration methods that differ from traditional ALT. Instead of conventional approaches, use-rate acceleration can be achieved by running algorithms at higher utilization rates. Another form of acceleration is input-data acceleration, such as error injection (EI) in \shortciteN{Panetal2024} and adversarial attacks in \shortciteN{Faddietal2024}. Thus, the concept of ALT can be valuable for AI testing.

\section{Datasets and Illustrations}\label{sec:datasets.illustration}

\begin{figure}
	\centering
	\includegraphics[width=1\textwidth]{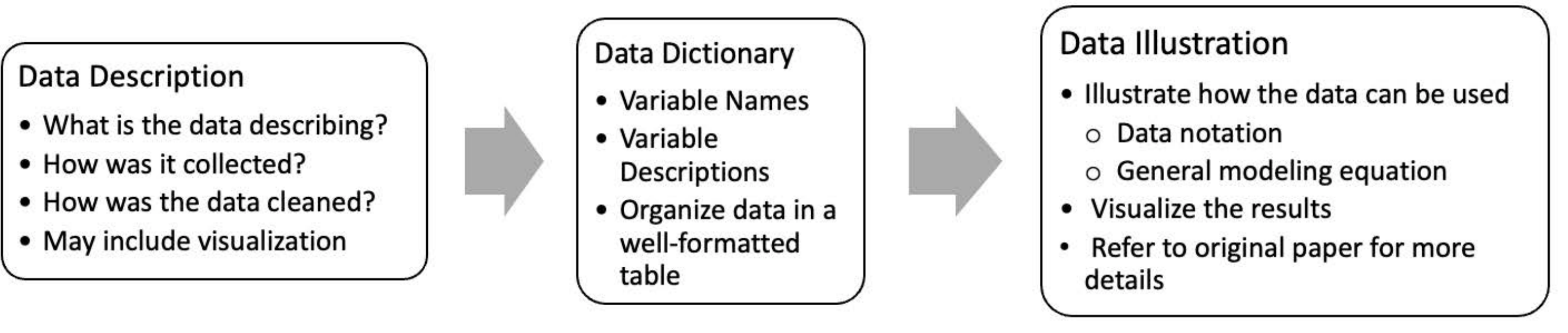}
	\caption{Flowchart for dataset introduction and illustrations.}\label{fig:flow}
\end{figure}

Now, we introduce the datasets we have collected and illustrate their applications in reliability modeling and analysis. To ensure a consistent presentation, Figure~\ref{fig:flow} outlines the flowchart for introducing the available datasets. Each dataset in Section~\ref{sec:datasets.illustration} will be presented according to this structure, starting with the data description, followed by the data dictionary, and then moving on to the data illustration.

\subsection{General AI Incidence Data}

\subsubsection{Data Description}

The website \citeANP{AIIncidentDB} database~(2024) documents incidents involving the use of AI systems that result in harm or near-harm consequences. 878 incidents have been reported. The reports are in text format, requiring substantial effort in data cleaning before the entries can be used for analysis. \shortciteN{hong2023statistical} cleaned up the data entries up until October 09, 2021.

After manually cleaning each entry, 72 reliability-related incidents were identified out of the 126 total incidents analyzed in \shortciteN{hong2023statistical}. Notably, 29 incidents out of the 72 events involve deaths or injuries, highlighting the importance of studying reliability issues.

\subsubsection{Data Dictionary}

The study by \shortciteN{hong2023statistical} then derived several variables from the text narratives in the original data entries to facilitate further analysis.
Table~\ref{tab:ai.incidence.data} shows the variables in the cleaned dataset.  These variables were carefully designed to capture key aspects of the incidents, enabling a structured and systematic examination of the data.

\begin{table}
\caption{Data dictionary for the AI incident database.}\label{tab:ai.incidence.data}
{\small
\begin{center}
\begin{tabular}{l|l} \hline\hline
Variable & Description\\\hline
IncidentNo & Incident case number.\\\hline 	
Company & Company for the system.\\\hline 	
Sector & Sector of the company.\\\hline		
System &  AI system. \\\hline 	
Algorithm & Algorithm(s) used in the system. \\\hline  	
Cause & Cause of the incident. \\\hline	
IncidentDescription & Description of the incident.\\\hline 	
Casuality & Is casuality established? \\\hline
Injured & Any human injured?	\\\hline
Comment & Additional comments for the incident. \\\hline\hline
\end{tabular}
\end{center}
}
\end{table}

\subsubsection{Data Illustration}

\begin{figure}
\centering
\begin{tabular}{cc}
\includegraphics[width=.45\textwidth]{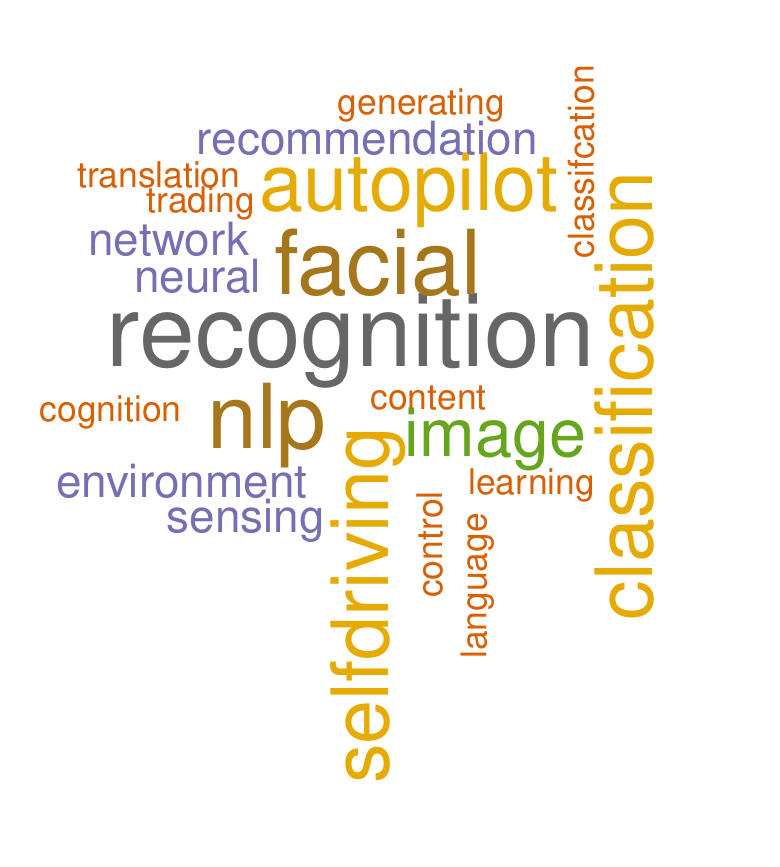}&
\includegraphics[width=.45\textwidth]{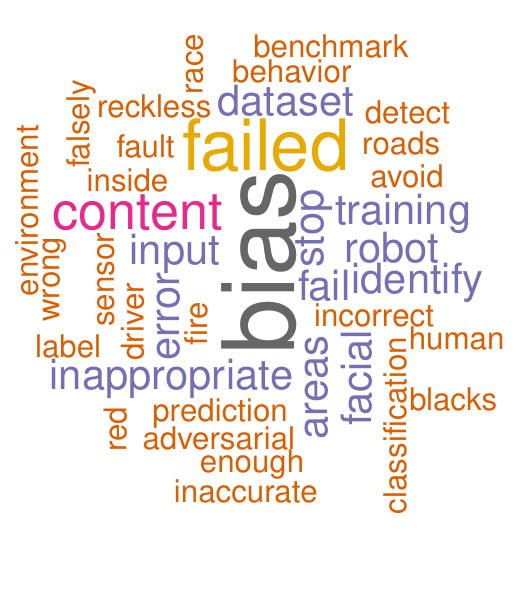}\\
(a) Algorithms Used in AI System & (b) Causes of Failures\\
\end{tabular}
\caption{Illustrations of AI system's algorithms and failure causes.}\label{fig:algo.cause.cloud.plot}
\end{figure}

 As an illustration of how the dataset can be used, Figure~\ref{fig:algo.cause.cloud.plot}(a) presents a word cloud that visualizes the different types of algorithms mentioned in the data entries. It shows that pattern recognition, self-driving systems, and NLP are among the most commonly used algorithms. Figure~\ref{fig:algo.cause.cloud.plot}(b) provides a word cloud that visualizes the causes of failure in these incidents, revealing that bias, inaccuracy, prediction errors, and adversarial attacks are key factors contributing to the failures.

The AI incident data can provide valuable insights into the causes of failures, but it cannot be used to infer the probability of an incident occurring. This is because the total number of deployed systems is unknown, and not all incidents may be reported. These are crucial considerations to keep in mind when interpreting the results of any analysis based on the AI incident data.

\subsection{Algorithm Level Test Data Set 1}
\subsubsection{Data Description}
\shortciteN{Lianetal2021Robustness} generated a test dataset to assess the robustness of AI classification algorithms, examining their performance quality and stability under class imbalance and distribution shifts between training and test datasets. The algorithms under investigation were XGboost used in \shortciteN{chen2015xgboost} and CNN used in \shortciteN{kim2014convolutional}. The dataset originates from carefully controlled experimental runs of two classification algorithms applied to two datasets: the KEGG dataset, which provides pathway data from the Kyoto Encyclopedia of Genes and Genomes, and the Bone Marrow dataset, which features macrophage scRNA-seq data.

Both datasets initially contain three distinct class labels in balanced proportions. To introduce class imbalance in the training and test datasets, the authors resampled the three classes from the original datasets. The data was collected in a structured format, defining the class proportions as $x_1, x_2,$ and $x_3$, the AI algorithm as $z_1$, and the dataset source (training or test) as $z_2$. Class imbalance was introduced by adjusting the proportions of the three classes using an adjusted centroid design (\citeNP{cornell2011experiments}). To simulate distribution shifts between training and test datasets, \shortciteN{Lianetal2021Robustness} considered balanced, consistent, and reverse scenarios. Model performance is measured with two key metrics: the mean AUC across classes and the log of the standard deviation of AUC values, which are computed as,
\begin{align*}
y_1 = \bar{\eta} = \frac{1}{m}\sum_{j=1}^{m} \eta_{j} \quad \text{and}\quad
y_2 = \log\left(\left[ \frac{1}{m-1} \sum_{j=1}^{m} (\eta_{j} - \bar{\eta})^{2}\right]^{1/2}\right),
\end{align*}
where $\eta_{j}$ is the AUC score of each class.
These metrics served to quantify both the accuracy and robustness of classification performance.

\subsubsection{Data Dictionary}

The details of the variables are displayed in Table~\ref{tab:data_dictionary}. For each combination of variable configurations, the experiment was repeated three times to collect the data, ending in $252$ total experimental observations collected. The type of responses are continuous variables.

\begin{table}
\caption{Data dictionary for evaluating the robustness of the AI classification algorithm.}
{\small
\begin{center}
\begin{tabular}{l|l} \hline\hline
Variable & Description \\ \hline
\texttt{$x_1$} & Proportion of class $1$ in the training dataset.  \\\hline 	
\texttt{$x_2$} & Proportion of class $2$ in the training dataset.  \\\hline 	
\texttt{$x_3$} & Proportion of class $3$ in the training dataset. \\\hline 	
\texttt{$z_1$} & Is the XGBoost algorithm applied? \\\hline 	
\texttt{$z_2$} & Is the KEGG dataset used?  \\ \hline
\texttt{$c_1$} & Is the experiment conducted under a balanced scenario?  \\ \hline
\texttt{$c_2$} & Is the experiment conducted under a consistent scenario?  \\ \hline
\texttt{$c_3$} & Is the experiment conducted under a reverse scenario?   \\ \hline
\texttt{$y_1$} & Mean AUC across the three classes. \\ \hline
\texttt{$y_2$} & Logarithm of standard deviation of AUC.  \\ \hline\hline
\end{tabular} \label{tab:data_dictionary}
\end{center}
}
\end{table}

\subsubsection{Data Illustration}

\begin{figure}
\centering
\begin{tabular}{cc}
\includegraphics[width=.25\textwidth]{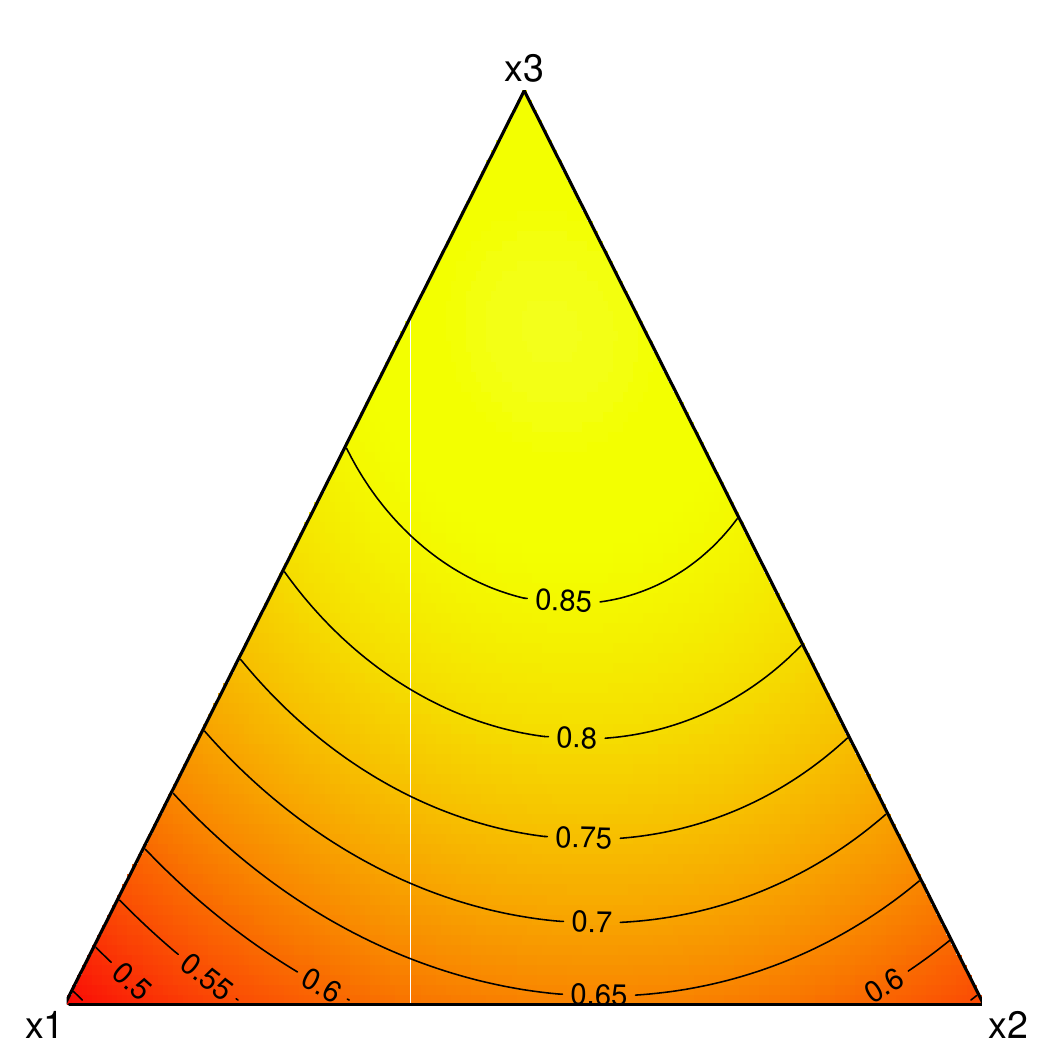}&
\includegraphics[width=.25\textwidth]{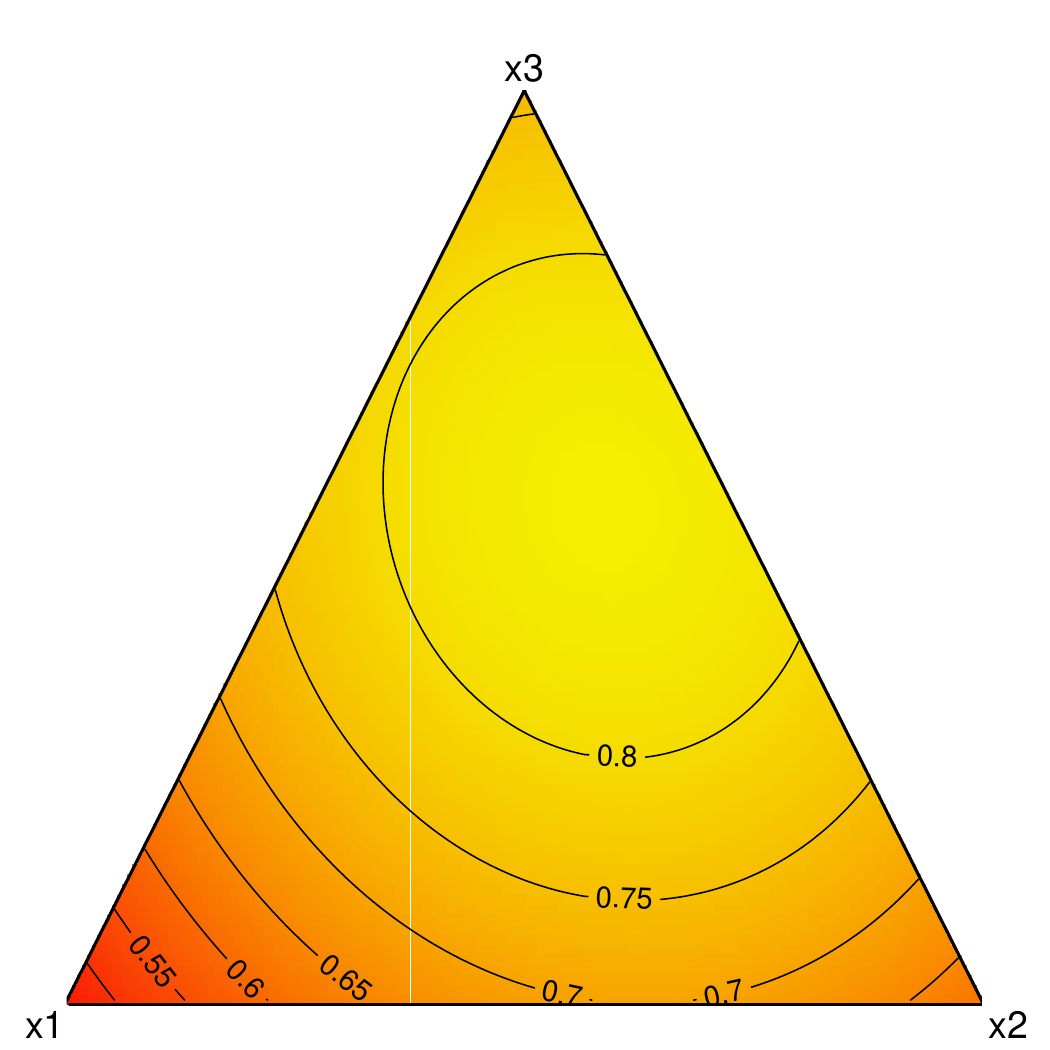}\\
(a) CNN with Bone Marrow  & (b) CNN with KEGG \\
\includegraphics[width=.25\textwidth]{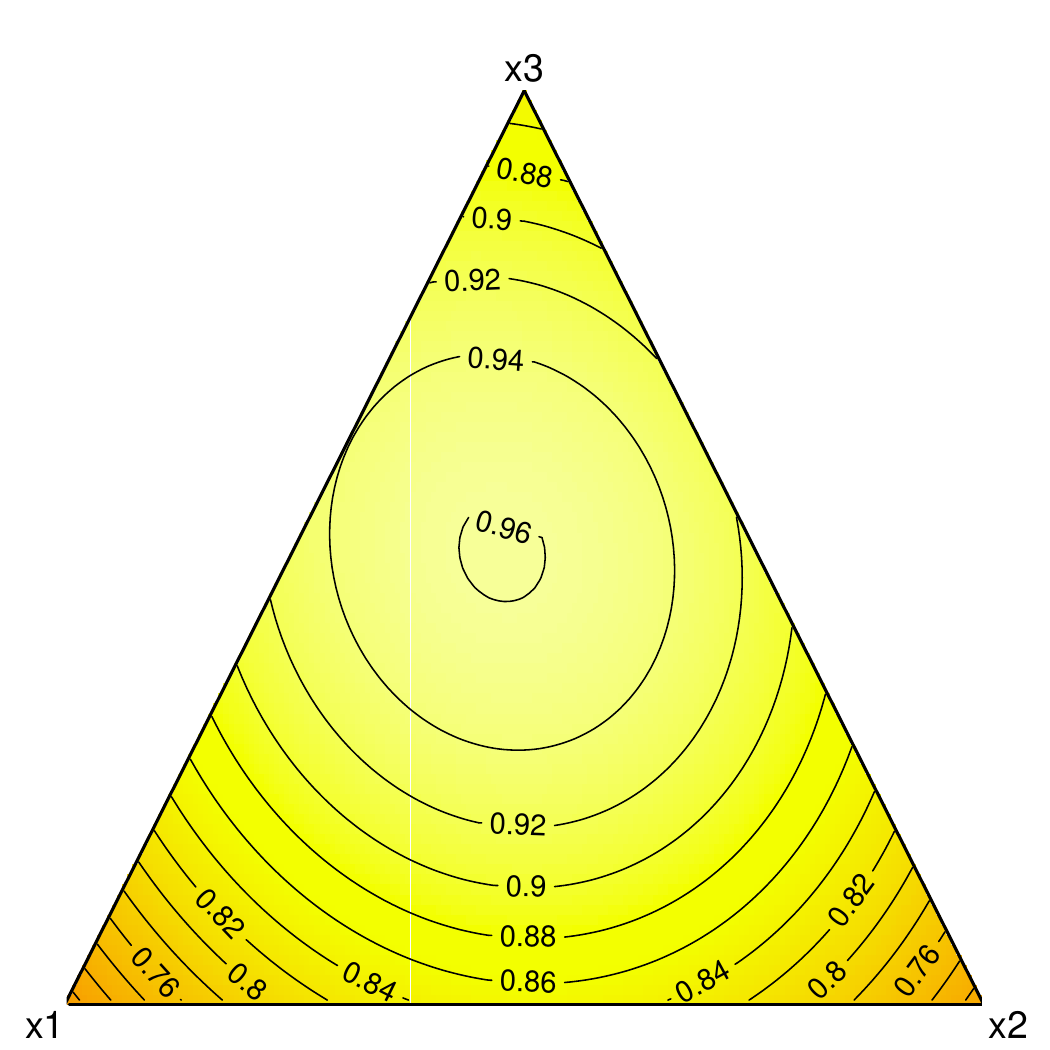}&
\includegraphics[width=.25\textwidth]{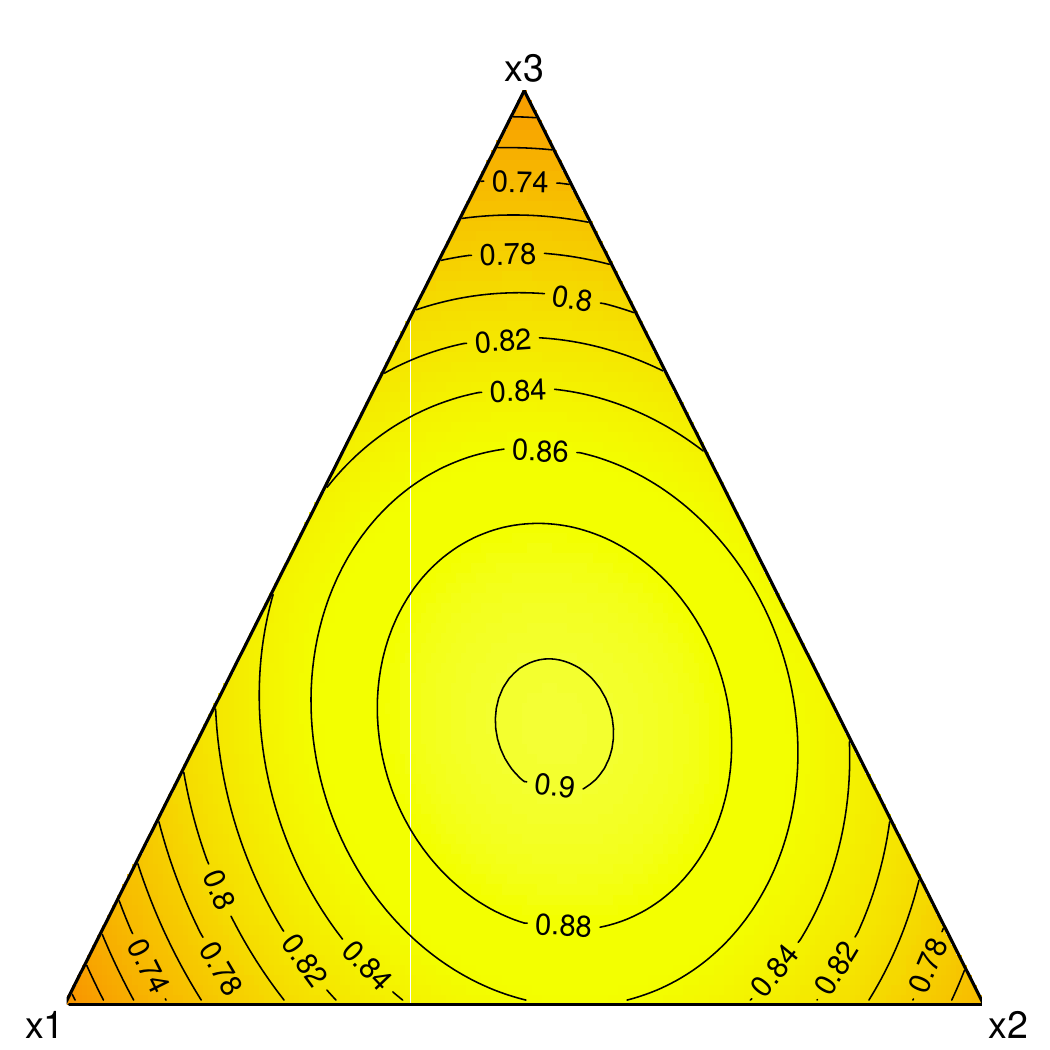}\\
(c) XGboost with Bone Marrow & (d) XGboost with KEGG
\end{tabular}
\caption{Contour plots of the predicted mean AUC across four different settings. \emph{Figure reproduced with permission from Taylor \& Francis.}}\label{fig:contour-AUC}
\end{figure}

To model the dataset, a regression model is employed that accounts for both main effects and interactions among predictors and covariates, commonly used in mixture design modeling. The model is formulated as follows:
\begin{align}\label{eqn:reg.model}
y &= \sum_{j=1}^{m} \beta_j x_{j} +  \sum_{j < j'} \beta_{j j'}x_{j}x_{j'} + \sum_{k=1}^{h} \sum_{j=1}^{m} \gamma_{kj} z_{k} x_{j} + \sum_{k < k'}\delta_{kk'} z_{k}z_{k'} +\epsilon,
\end{align}
where $m=3$, $h=2$, and $\beta_j, \beta_{j j'}$, $\gamma_{kj}$ and $\delta_{kk'}$ are regression coefficients.  Figure~\ref{fig:contour-AUC} displays triangular contour plots depicting the predicted mean AUC under the balanced scenario. Overall, training with balanced datasets leads to higher accuracy. For the Bone Marrow dataset, both algorithms require a higher proportion of $x_3$ to achieve the maximum response value. XGBoost is better than CNN across both datasets. Additionally, CNN prioritizes $x_3$ more strongly, whereas XGBoost exhibits a more systematic response pattern.

\subsection{Algorithm Level Test Data Set 2}
\subsubsection{Data Description}

\shortciteN{Faddietal2024} presents a dataset that investigates the performance of CNNs on both clean and perturbed inputs. This dataset serves as a foundation for assessing the reliability and resilience of image recognition systems under adversarial conditions. To evaluate the reliability and resilience of CNNs, experiments were conducted on an image recognition system to capture the behavior of an ML classifier on clean and perturbed inputs, enabling performance analysis across iterative retraining cycles. Initially, the classifier was trained on a subset of the publicly available CIFAR-10 dataset~\cite{krizhevsky2009learning} to learn patterns in the data. CIFAR-10 comprises 60{,}000 color images of size $32 \times 32$. Those images are categorized into ten classes, with each class containing 6,000 images. Later, the classifier was tested on different datasets containing fake images generated by various adversarial attacks, which aimed to manipulate the ML algorithm with malicious inputs, leading to incorrect predictions or degraded system performance.

\begin{figure}
\begin{center}
\includegraphics[width=0.65\textwidth]{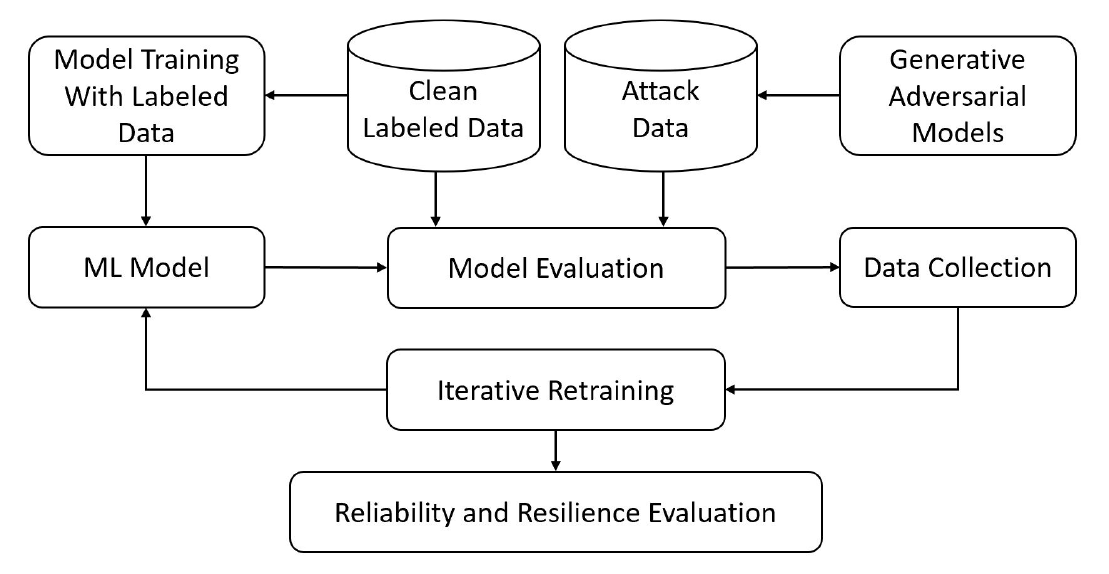}
\caption{Data collection process to assess the performance of CNN.}\label{fig:framework} 
\end{center}
\end{figure}

Figure~\ref{fig:framework}  illustrates the data collection process, which involves the following steps. First, one trains the CNN model, using random $50{,}000$ clean labeled images from the CIFAR-10 dataset until achieving a specified performance threshold (e.g., an initial 70\% accuracy) required for deployment. Second, adversarial examples are generated using the remaining $10{,}000$ images from the CIFAR-10 dataset with the Fast Gradient Sign Method (FGSM) and the Projected Gradient Descent (PGD) method, applying noise levels sampled from a uniform distribution ($0 \leq \varepsilon \leq 1$), where $\varepsilon$ represents the perturbation magnitude. Third, clean and perturbed data are combined to create a poisoned dataset and evaluate model performance against adversarial attacks. Failure metrics (e.g., misclassifications, accuracy, and loss) and test performance were recorded. Fourth, the model is retrained with poisoned data to improve reliability and resilience, repeating adversarial exposure over multiple iterations (e.g., 30).

\subsubsection{Data Dictionary}

Table~\ref{tab:ML.algorithm} provides a detailed description of the variables collected for the failure count dataset during the training and evaluation of the CNN, categorized into pre-retraining and post-retraining metrics. An additional dataset records the failure time, using the index of the misclassified image as the failure time. The code and data are available in a public GitHub repository in \citeN{RRML2025}.

\begin{table}
\caption{Data dictionary to evaluate the reliability and resilience of AI algorithms.}\label{tab:ML.algorithm}
{\small
\begin{center}
\begin{tabular}{l|l} \hline\hline
Variable & Description\\\hline	
Scenario & Different scenarios correspond to different epsilon ranges.\\\hline
EpsilonRange & The epsilon range for a specific scenario.\\\hline
T & The number of steps.\\\hline		
FC & Failure Count, also denoted as FN. \\\hline 	
Alpha & Learning rate used during retraining \\\hline  	
F1 & F1-Score computed for the model on the poisoned dataset. \\\hline	
Epsilon & Magnitude of noise applied to input samples. \\\hline 	
FGSM & The percentage of the $5000$ adversarial attacks using FGSM.\\\hline
PGD & The percentage of the $5000$ adversarial attacks using PGD. \\\hline
TrainingAccuracy & Accuracy of the model following the retraining step.\\\hline
TrainingLoss & Loss of the model following the retraining step.\\\hline
ValidationAccuracy  & Accuracy of the model following the retraining step.	\\\hline
ValidationLoss  & Loss of the model following the retraining step.\\\hline
TestAccuracy  & Accuracy of the model on the poisoned dataset.	\\\hline
TestLoss  &  Loss recorded for the model on the poisoned dataset.	\\\hline
Memory & Memory consumption during iterative retraining.\\\hline\hline
\end{tabular}
\end{center}
}
\end{table}

\subsubsection{Data Illustration}

As an illustration, we briefly describe the modeling and analysis conducted in \shortciteN{Faddietal2024}.  The grouped failure count is used as the response variable for the reliability models, and test accuracy is used as the response variable for the resilience models. The remaining factors collected were treated as covariates.

First, for software reliability, software reliability growth models, which may incorporate covariates, are commonly used to estimate reliability metrics (\shortciteNP{2020nagaraju}, and \shortciteNP{2006shibata}). These models provide a mean value function $m(t; \xvec)$, which predicts the cumulative number of failures discovered up to time interval $t$, given covariates $\xvec_s$. The mean value function is defined as:
\begin{equation} \label{eq:cox-MVF}
m(t; \xvec) = \omega \sum_{l=1}^{t} \big((1-(1-h(l))^{g(\xvec_l;\betavec)})\prod\limits_{s=1}^{l-1}
(1-h(s))^{g(\xvec_s;\betavec)}\big),
\end{equation}
where $\omega > 0$ represents the total number of failures that would be observed with infinite testing, $h(\cdot)$ is the baseline hazard function, $g(\xvec_{l}; \betavec)$ is a general function of covariates $\xvec_l$ and parameter vector $\betavec$, capturing the impact of external factors on software reliability, $l$ represents the current time interval at which failures are being counted, and $s$ is an index for prior time intervals.

Specifically, the geometric model (GM), negative binomial of order two (NB2), discrete Weibull of order two (DW2), type III discrete Weibull (DW3), S distribution (S), and truncated logistic (TL) can be used to model the baseline hazard function $h(\cdot)$. The covariates can be modeled as follows:
\begin{equation} \label{eq:g-function}
g(\xvec_{t};\betavec) = \exp(\beta_{1} x_{t1} + \beta_{2}x_{t2} + \dots + \beta_{m}x_{tm}),
\end{equation}
where $\xvec_{t}=(x_{t1}, \dots,x_{tm})'$ is a vector of $m$ covariates at time $t$.

\begin{figure}
\begin{center}
\begin{tabular}{cc}
\includegraphics[width=0.48\textwidth]{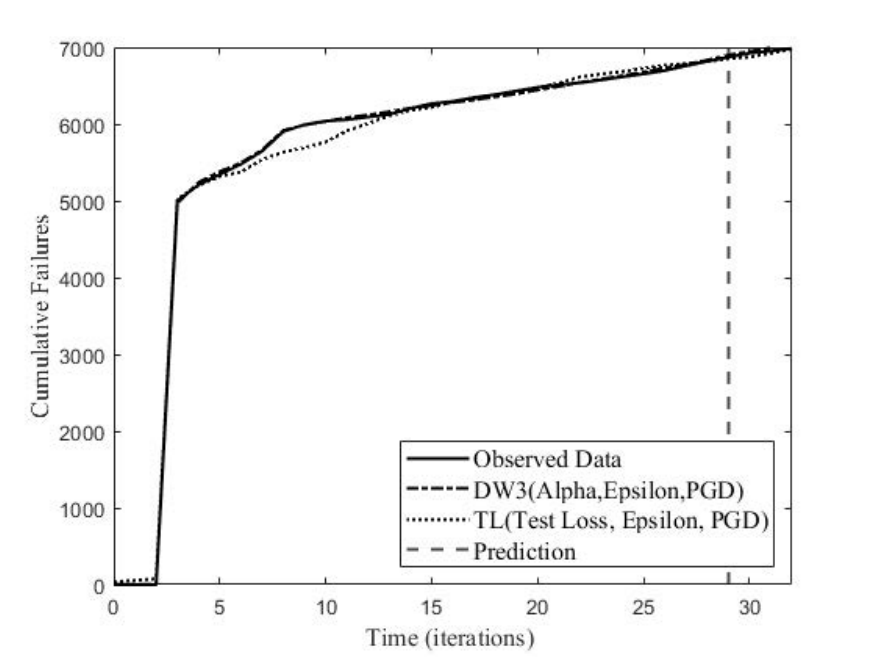}&
\includegraphics[width=0.48\textwidth]{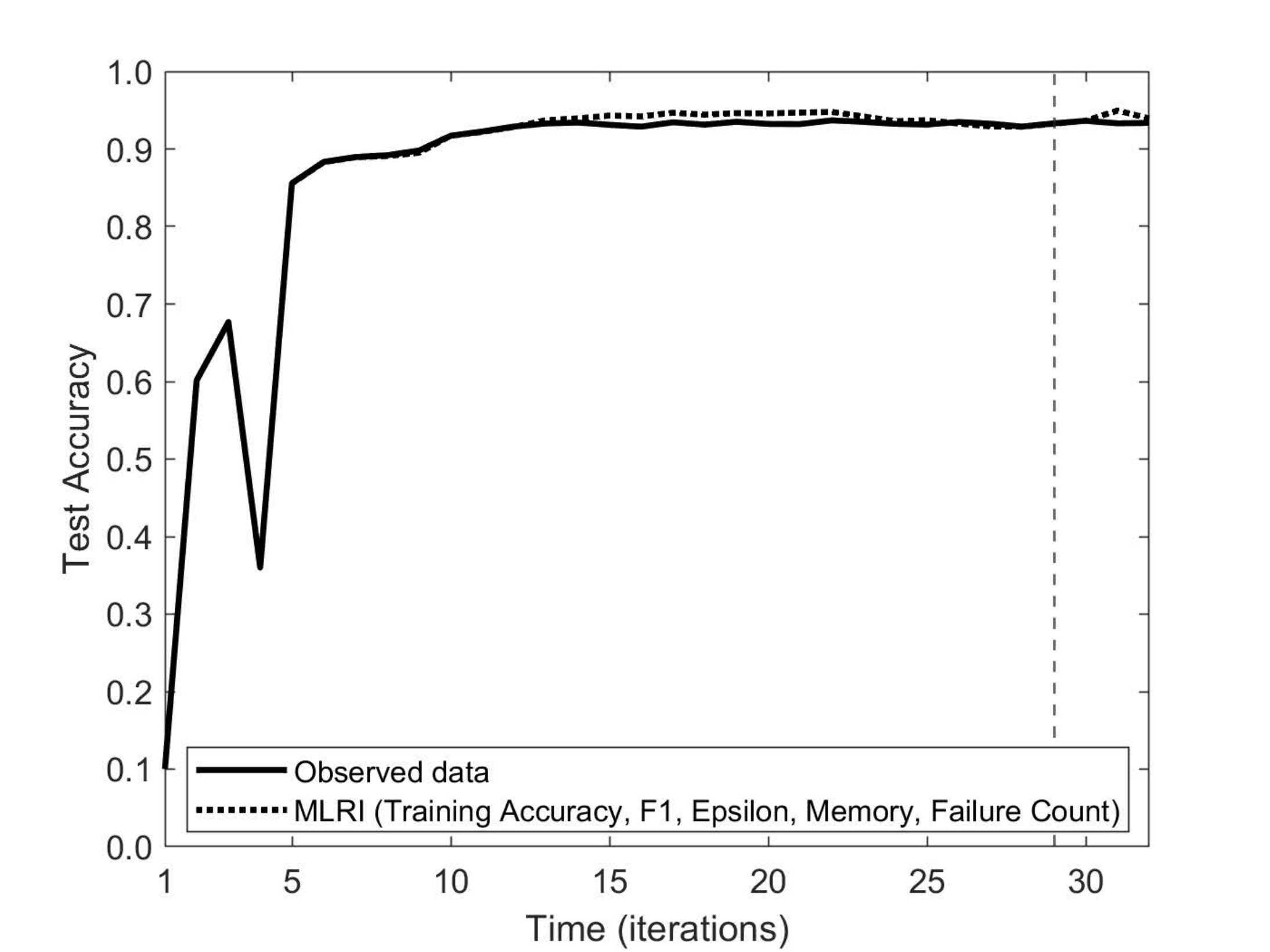}\\
(a) Reliability  & (b) Resilience
\end{tabular}
\caption{Observed cumulative failure counts and best model fit (a) and observed cumulative failure counts and best model fit (b). }
\label{fig:bestmodelReliability.resilience}
\end{center}
\end{figure}

The optimal subset of covariates for each hazard function was selected using forward stepwise selection, applying maximum likelihood estimation with $90\%$ of the dataset to estimate the parameters of each model and predict the rest $10\%$ not used for model fitting. After fitting various mean value functions with different hazard functions, Figure~\ref{fig:bestmodelReliability.resilience}(a) presents the two best-fitting covariate models, incorporating the DW3 and TL hazard functions along with their respective optimal sets of covariates.

Then, resilience models can characterize the decreases and increases in the performance of a system as a function of the intensity of disruptive events and restorative efforts (\shortciteNP{Silva2024}). To model resilience, let $r(t)$ represent the performance in the present interval and $r(t-1)$ represent the performance in the previous interval. We define their relationship as follows:
$$r(t)=r(t-1)+\Delta r(t),$$
where $\Delta r(t)$ denotes the change in performance. More specifically, to model $\Delta r(t)$, one can use regression models such as multiple linear regression and polynomial regression. For example, with linear regression,
\begin{equation}\label{eq:MLR}
    \Delta {r}(t)=\beta_0 + \sum_{j=1}^{m}\beta_j x_j(t),
\end{equation}
where $\beta_0$ represents the baseline change in performance, $x_j(t)$ denotes the detrimental or restorative covariates, and $\beta_j$ are their corresponding coefficients, characterizing the impact of hazards or efforts on performance, with $j=1,\dots, m$.

For model selection and estimation of the three aforementioned regression-based resilience models, one can use the stepwise selection method to identify the optimal set of covariates for each model. Figure~\ref{fig:bestmodelReliability.resilience}(b) illustrates the best-fitting resilience model, identified as multiple linear regression with interaction, along with its corresponding optimal covariates. Figure~\ref{fig:bestmodelReliability.resilience}(b) illustrates how the accuracy of the model initially drops due to adversarial attacks but recovers and improves after the implementation of adaptive adversarial training. For more details, we refer to \shortciteN{Faddietal2024}.

\subsection{Module Level Test Data}
\subsubsection{Data Description}
\shortciteN{Panetal2024} introduced a dataset containing module-level error events from AI systems in AVs operating across various driving scenarios. The tested modules belong to the perception system, which comprises cameras and LiDAR sensors. This system includes three key modules: 2-dimensional (2-D) detection, 3-D detection, and object localization. The 2-D and 3-D detection modules operate in parallel, and their outputs are fused in the localization module to determine object positions.

\begin{figure}
	\centering
	\includegraphics[width=.75\textwidth]{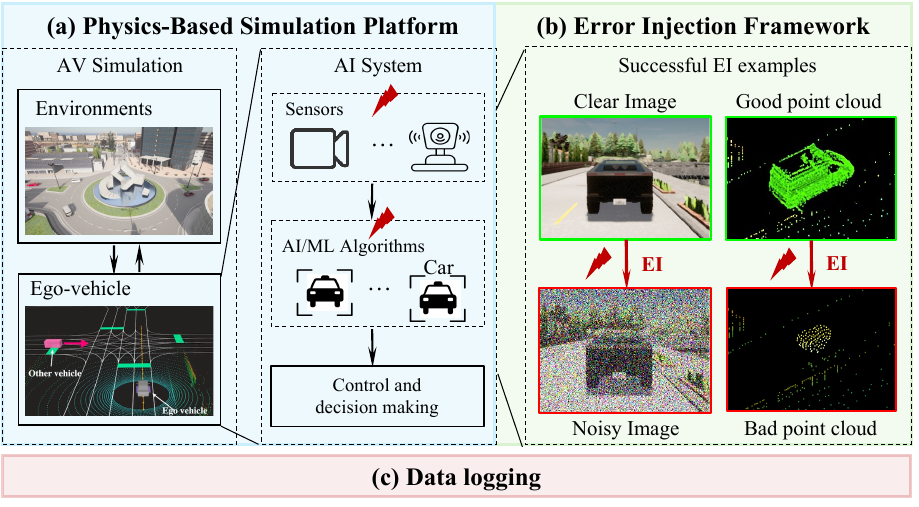}
	\caption{EI framework for testing AV in a physics-based simulation platform.}\label{fig:Simulation_framework}
\end{figure}

The dataset was collected from a physics-based AV simulation platform, where an EI framework was developed to efficiently generate error events from various AI system modules in AVs, as shown in Figure~\ref{fig:Simulation_framework}. Figure~\ref{fig:Simulation_framework}(a) illustrates the physics-based simulation platform, which consists of two main components: (i) the environment, incorporating diverse physical models such as infrastructures, driving scenarios, and traffic-related agents that closely resemble real-world driving conditions, and (ii) the ego vehicle, which interacts with the driving environment through an AI system that integrates multiple sensors and AI/ML algorithms to perceive environmental information. Figure~\ref{fig:Simulation_framework}(b) depicts the EI framework, which enables targeted EI into different AI system modules at user-defined time stamps and probabilities. Recurrent error events were logged throughout the simulation process, as shown in Figure~\ref{fig:Simulation_framework}(c). The primary objective of the dataset is to analyze how errors in the 2-D and 3-D detection modules propagate to the object localization module.

\subsubsection{Data Dictionary}
Table~\ref{tab:module-levelr} presents the data dictionary for module-level error events in the AI system of AVs. Seven scenarios were considered, with EI controlled by the timing parameter for module $m$, $t^{\err}_{m}$, and probability, $p^{t^{\err}}_m$. Each scenario was simulated for 20 seconds. The weather conditions included persistent clear, snowy, rainy, and foggy, as well as intermittent snowy, rainy, and foggy. In one setting, errors were injected throughout the entire interval ($t^{\err}_{m} \in [0, 20)$), while in another setting, errors were injected only during the second half of the interval ($ t^{\err}_{m} \in [10, 20)$).

\begin{table}
	\caption{Data dictionary for the module-level error events from AI system of AVs.}
	\label{tab:module-levelr}
{\small
\begin{center}
\begin{tabular}{l|l} \hline\hline
Variable & Description \\ \hline
ScenarioID & Identifier for each of the seven simulated driving scenarios. \\ \hline
Weather & Simulated weather conditions during the driving scenario. \\ \hline
Observation window & Time interval (in seconds) for observing and recording events. \\ \hline
EI time in 2D module & Time interval for error injected into the 2D module. \\ \hline
EI prob in 2D  module & Probability for injecting errors into the 2D detection module. \\ \hline
EI time in 3D module & Time interval for error injected into the 3D  module. \\ \hline
EI prob in 3D module & Probability for injecting errors into the 3D detection module. \\ \hline
TimeStamp & Time at which an error event occurred or error free. \\ \hline
2D error indicator & Indicates whether a 2D miss detection error occurred. \\ \hline
3D error indicator & Indicates whether a 3D miss detection error occurred. \\ \hline
Localization error indicator & Indicates whether a miss localization error occurred. \\ \hline\hline
\end{tabular}
\end{center}
}
\end{table}

\subsubsection{Data Illustration}
\shortciteN{Panetal2024} proposed an error propagation (EP) model to describe the recurrent error events data, which is based on NHPP. For module $m$, let $N_m([t_1, t_2))$ be the counting process that records the number of events that occurred in time interval $[t_1, t_2)$. Given history $\Hs_m(t)$, the event intensity $\lambda_m(t)$ is defined as,
\begin{equation}\label{eq:lambda_nhpp}
\lambda_m(t|\Hs_m(t))=\lim_{dt \to 0} {\E[N_m([t, t+dt))|\Hs_m(t)]}/{dt}.
\end{equation}
Let $\Lambda_m(t_1, t_2) = \int_{t_1}^{t_2}\lambda_m(u)du$ be the cumulative intensity function (CIF). The power-law function, $\lambda_m(t;\thetavec) = (\beta/\eta)(t/\eta)^{\beta-1}, \beta>0, \eta>0$ is widely used for event intensity. The parameters are denoted as $\thetavec = (\beta, \eta)'$.

As defined in \eqref{eq:lambda_nhpp}, the NHPP is able to model the intensity function for each individual module without considering the impact from other modules. To model the EP between different modules, an event-triggering point process was proposed in \shortciteN{pan2022quantifying} and \shortciteN{Panetal2024}, where the intensity function of each module $m$ can be decomposed into two terms, i.e.,
\begin{equation}
 \label{eq:lambda_hp}
    \underbrace{\lambda_m(t|\Hs_m(t), \Hs_1(t), \Hs_2(t), \cdots, \Hs_N(t))}_
    \text{\scriptsize Overall error intensity}
    = \underbrace{\lambda_{m}^0(t|\Hs_{m}(t))}_
    \text{\scriptsize Baseline intensity} +  \underbrace{\sum\nolimits_{n=1}^{N} \lambda_{m, n}^p(t|\Hs_n(t))}_{
    \text{\scriptsize Triggering intensity}},
\end{equation}
where the baseline intensity, $\lambda_{m}^0(t|\Hs_{m}(t))$, is used to model the error caused by module $m$ itself and the triggering intensity is used to model the error propagated from other interdependent modules $n$, where $n = 1, \ldots, N$. Here, $N$ is the total number of modules functionally interdependent with module $m$. The baseline intensity and the triggering intensity can be defined as various parametric forms.

\begin{figure}
	\centering
	\includegraphics[width=.95\textwidth]{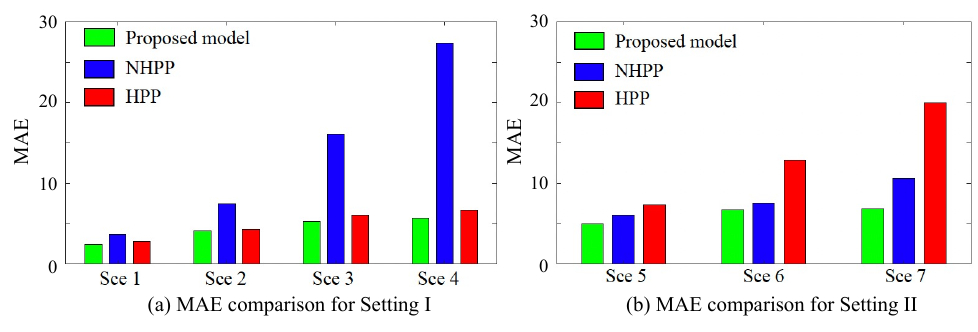}
	\caption{Performance comparison using different models for error AV error events data. \emph{Figure reproduced with permission from Elsevier Ltd.}}\label{fig:MAE}
\end{figure}

The log-likelihood for all modules in the system is:
\begin{equation}\label{eq:log-likelihood}
l(\thetavec|\text{Data}) = \sum_{m=1}^M\sum_{i=1}^{n_{m}}\log(\lambda_{m}(t_{mi})) - \int_0^{\tau} \lambda_{m}(t)dt,
\end{equation}
where $\lambda_{m}(t)$ is the intensity function of module $m$, and $\thetavec$ represents the parameter set. By accounting for EP between different modules, \shortciteN{Panetal2024} demonstrated that the event-triggering point process achieves superior reliability prediction performance, yielding a lower mean absolute error (MAE) compared to commonly used homogeneous Poisson process (HPP) and NHPP methods, as illustrated in Figure~\ref{fig:MAE}. Further details and results can be found in \shortciteN{Panetal2024}.

\subsection{System Level Test Data Set 1}\label{sec:disengagement}

\subsubsection{Data Description} %

The system-level test data analysis from \shortciteN{MinHongKingMeeker2020} focuses on the reliability of AVs. Disengagement event data is utilized to evaluate the reliability of AI systems. The original data is made available to the public by the California DMV. The data were collected through the Autonomous Vehicle Tester (AVT) program. In the AVT program, a human driver is required to sit in a test AV in order to take control of the vehicle when needed. Test AVs can disengage from the autonomous mode when the AI system or the human driver determines it is not safe to continue using the self-driving mode. Thus, the occurrence rate of disengagement events can be viewed as a representative of the reliability of the AI systems in the AVs. The original data contains exact dates of disengagement events for all the tested vehicles in the AVT program from December 2017. The monthly driven mileage information of the tested AVs is also available, allowing for more sophisticated reliability analysis of the AVs. The original data can be accessed from \citeN{CAdriving} and is updated yearly, as California DMV requires all the manufacturers who participate in the AVT program to report their disengagement events annually.

\begin{figure}
	\centering
	\begin{tabular}{cc}
		\includegraphics[width=.46\textwidth]{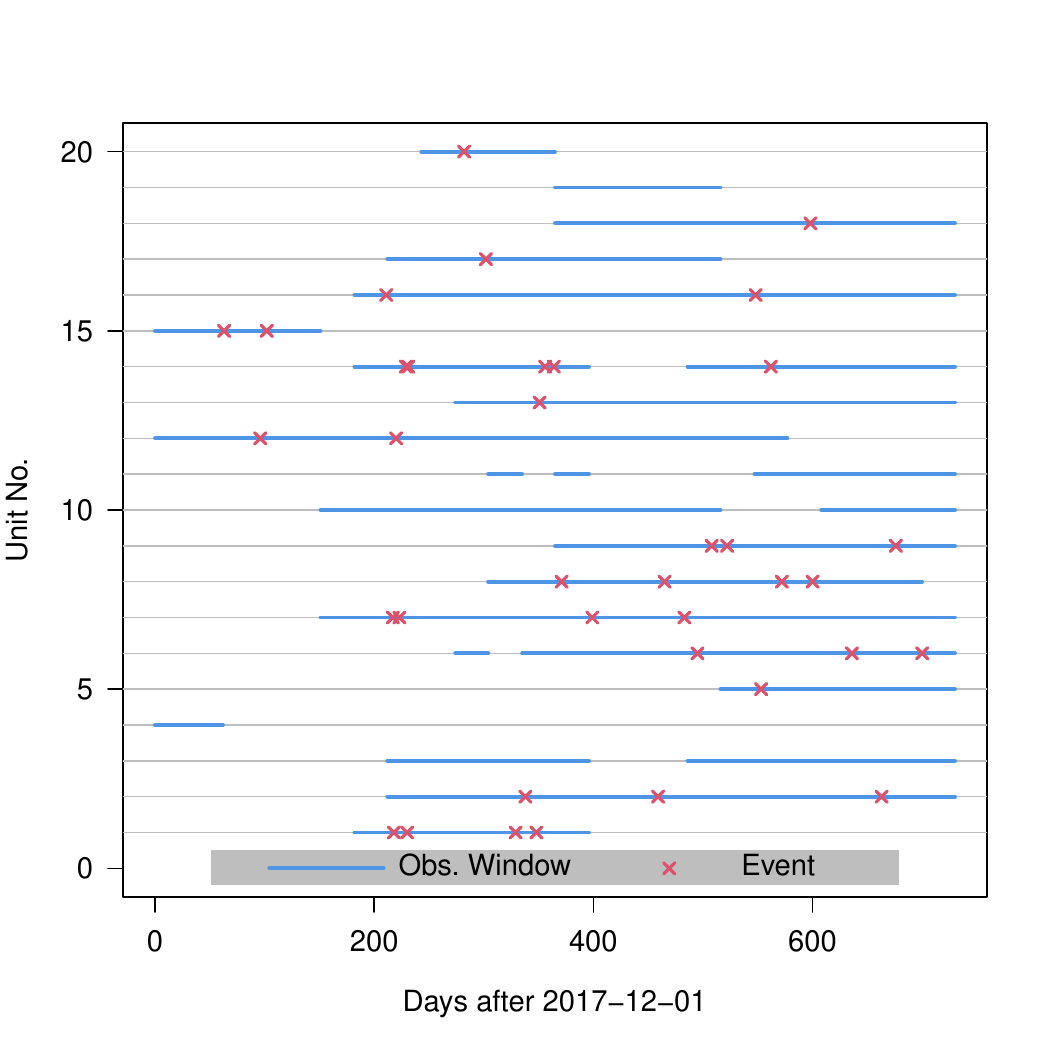}&
		\includegraphics[width=.46\textwidth]{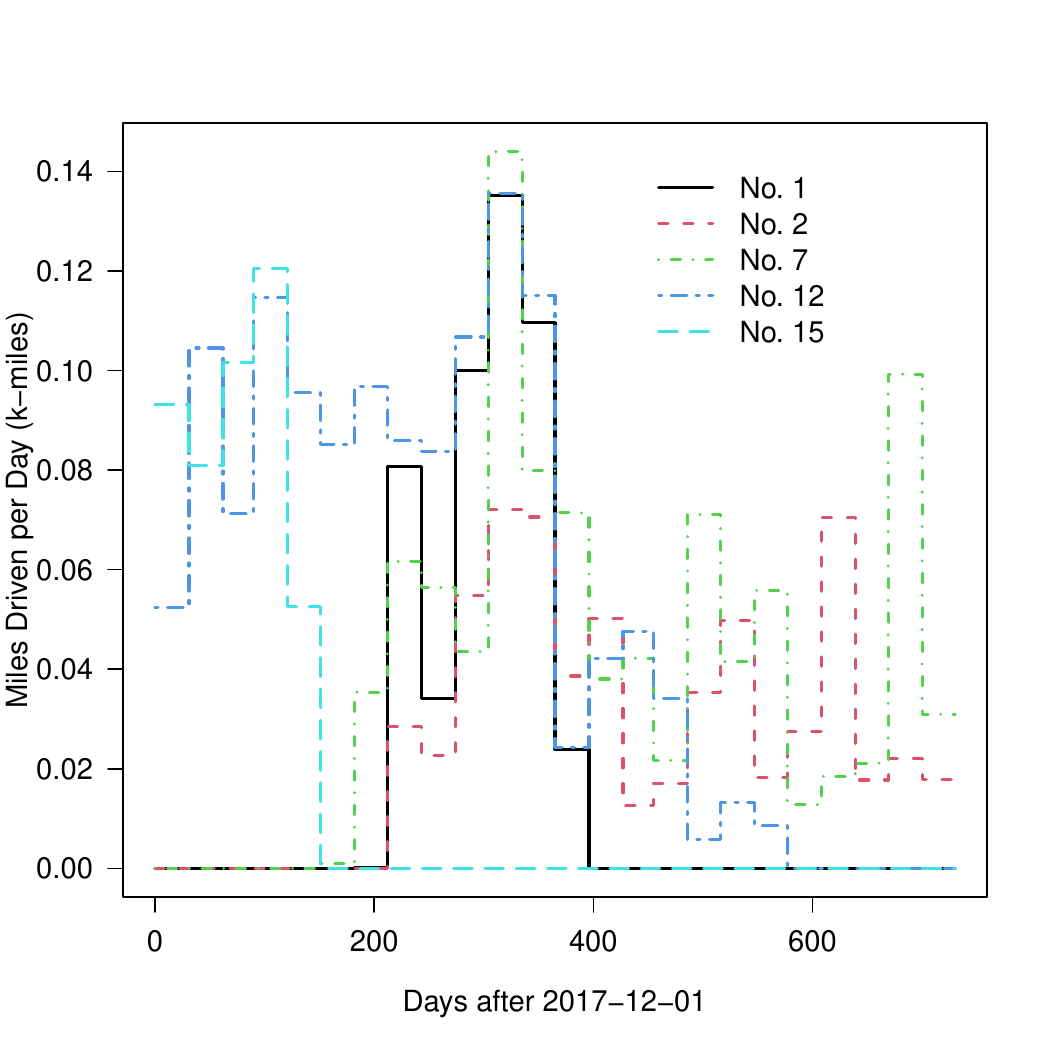}\\
		(a) Recurrent Events & (b) Thousands of Miles per day\\
	\end{tabular}
	\caption{Visualizations of a subset of disengagement data: (a) event times and observation windows for 20 vehicles, and (b) daily mileage for 5 vehicles. \emph{Figure reproduced with permission from Oxford University Press.}}\label{fig:dissample}
\end{figure}

\shortciteN{MinHongKingMeeker2020} cleaned the original disengagement data from December 1, 2017 to November 30, 2019, making it suitable to use in reliability analysis. The disengagement data provided in \shortciteN{MinHongKingMeeker2020} contains the disengagement events and related information reported from four manufacturers that performed extensive AV driving tests during the two year period: Waymo, Cruise, Pony AI, and Zoox. For the disengagement data provided in \shortciteN{MinHongKingMeeker2020}, the time scale for events is the number of days since the starting date (i.e., December 1, 2017). The unit for the monthly mileage is thousands of miles. Figure~\ref{fig:dissample}(a) shows the disengagement event times and observation windows for twenty vehicles from manufacturer Waymo.  Additionally, Figure~\ref{fig:dissample}(b) shows the daily mileage information of five tested vehicles from Waymo. The daily mileage is obtained by dividing the monthly mileage provided in the data by the number of days in that particular month.

\subsubsection{Data Dictionary}
The cleaned data provided by \shortciteN{MinHongKingMeeker2020} consists of three CSV files containing information on disengagement event times, mileage, and month information. Table~\ref{tab:dis.var} summarizes the variables related to disengagement event times. The mileage information file includes the variables manufacture and VIN, which can be used to link disengagement events with mileage data for each vehicle. Additionally, this file contains 24 numerical columns representing the monthly mileage for all vehicles over the 24-month period. The monthly information file provides details on the start date, end date, and the number of days in each of the 24 months, enabling the calculation of daily mileage for autonomous vehicles and supporting further reliability analysis. The data type for the response variable is recurrent events. While mileage can serve as a covariate in modeling, \shortciteN{MinHongKingMeeker2020} treated it as a measure of exposure.

\begin{table}
\caption{Data dictionary for the California DMV disengagement events dataset.}\label{tab:dis.var}
{\small
\begin{center}
\begin{tabular}{l|l} \hline\hline
Variable & Description\\\hline
Manufacture    & Manufacture of the AVs.  \\ \hline
VIN            & Unique vehicle identify number                                                                                   \\\hline
Date           & Disengagement event occurrence dates                                                                             \\\hline
Month         & Disengagement event occurrence months                                                                            \\\hline
MonthID        & Identify number for the 24 months in the 2-year period   \\      \hline\hline
\end{tabular}
\end{center}
}
\end{table}

\subsubsection{Data Illustration}
As an example of how the data can be utilized in reliability analysis, \shortciteN{MinHongKingMeeker2020} modeled the disengagement event processes using NHPP. Specifically, let $n$ represent the number of tested AVs, $\tau$ denote the duration of the testing period, and $t_{ij}$ be the time of event $j$ for unit $i$, where $i = 1, 2, \dots, n$ and $j = 1, 2, \dots, n_i$, with $n_i$ indicating the number of events for unit $i$ during the testing period. Additionally, let $x_i(t)$ represent the daily driven mileage for unit $i$ at time $t$, where $0 < t \leq \tau$. The intensity function for unit $i$ is
\begin{equation*}
	\lambda_i\left[t; \thetavec,x_i(t)\right] = \lambda_0(t; \thetavec)x_i(t),
\end{equation*}
where $\lambda_0(t; \thetavec)$ represents a common BIF shared by all $n$ units, $\thetavec$ contains the unknown parameters of the BIF, and $x_i(t)$ serves as an adjustment factor for the intensity function based on the vehicles' driven mileage. The CIF and cumulative baseline intensity function (CBIF) are then given by
\begin{align*}
\Lambda_i[t; x_i(t), \thetavec]=\int_{0}^{t}\lambda_0(s; \thetavec)x_i(s)ds & \;\textrm{and} \; \Lambda_0(t; \thetavec)=\int_{0}^{t}\lambda_0(s; \thetavec)ds,
\end{align*}
and the likelihood function for estimating $\thetavec$ is derived as
\begin{align}\label{eqn:event.lik}
L(\thetavec)=\prod_{i=1}^n\left\{\prod_{j=1}^{n_i} \lambda_i[t_{ij}; x_i(t_{ij}), \thetavec]\right\}\times \exp\{-\Lambda_i[\tau; \xvec_i(\tau), \thetavec]\}.
\end{align}

\begin{figure}
	\centering
	\begin{tabular}{cc}
		\includegraphics[width=.46\textwidth]{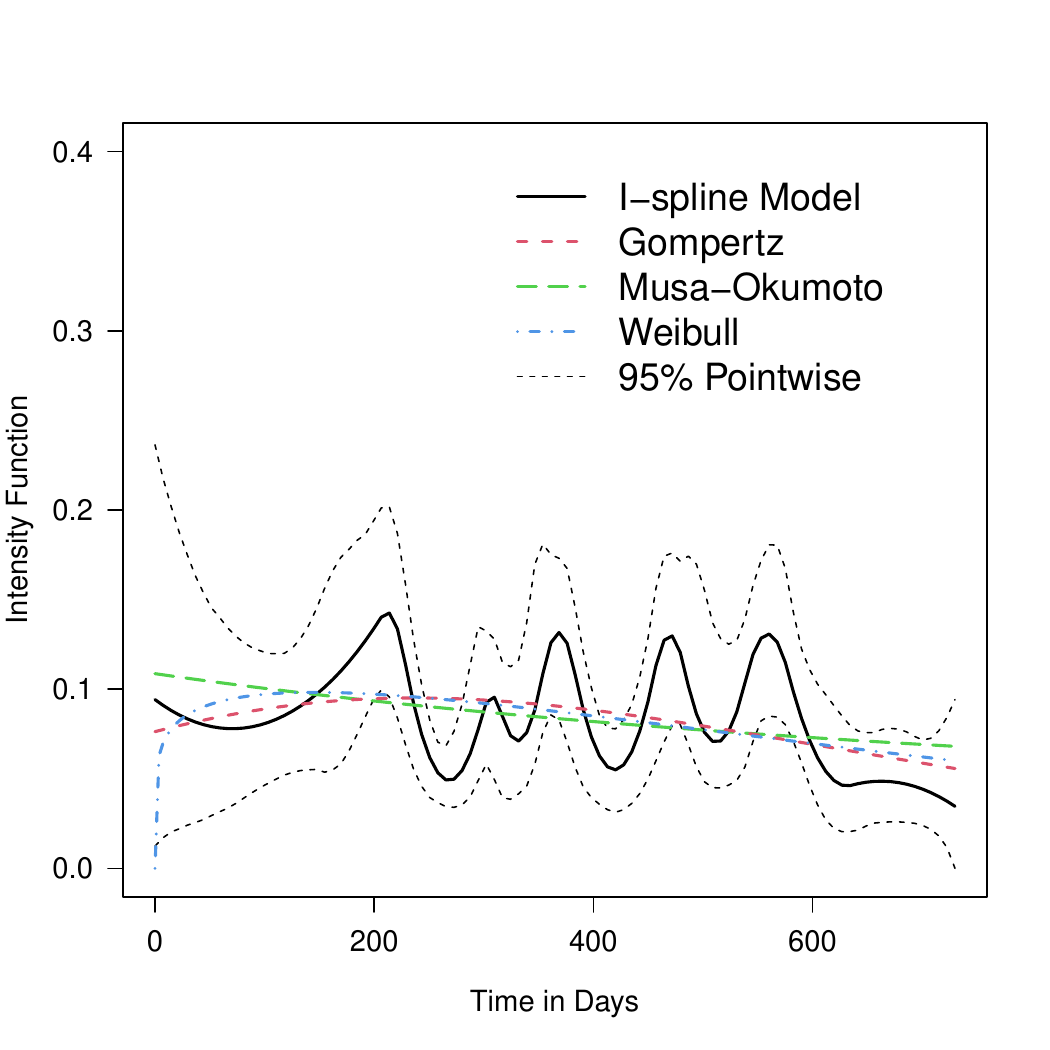}&
		\includegraphics[width=.46\textwidth]{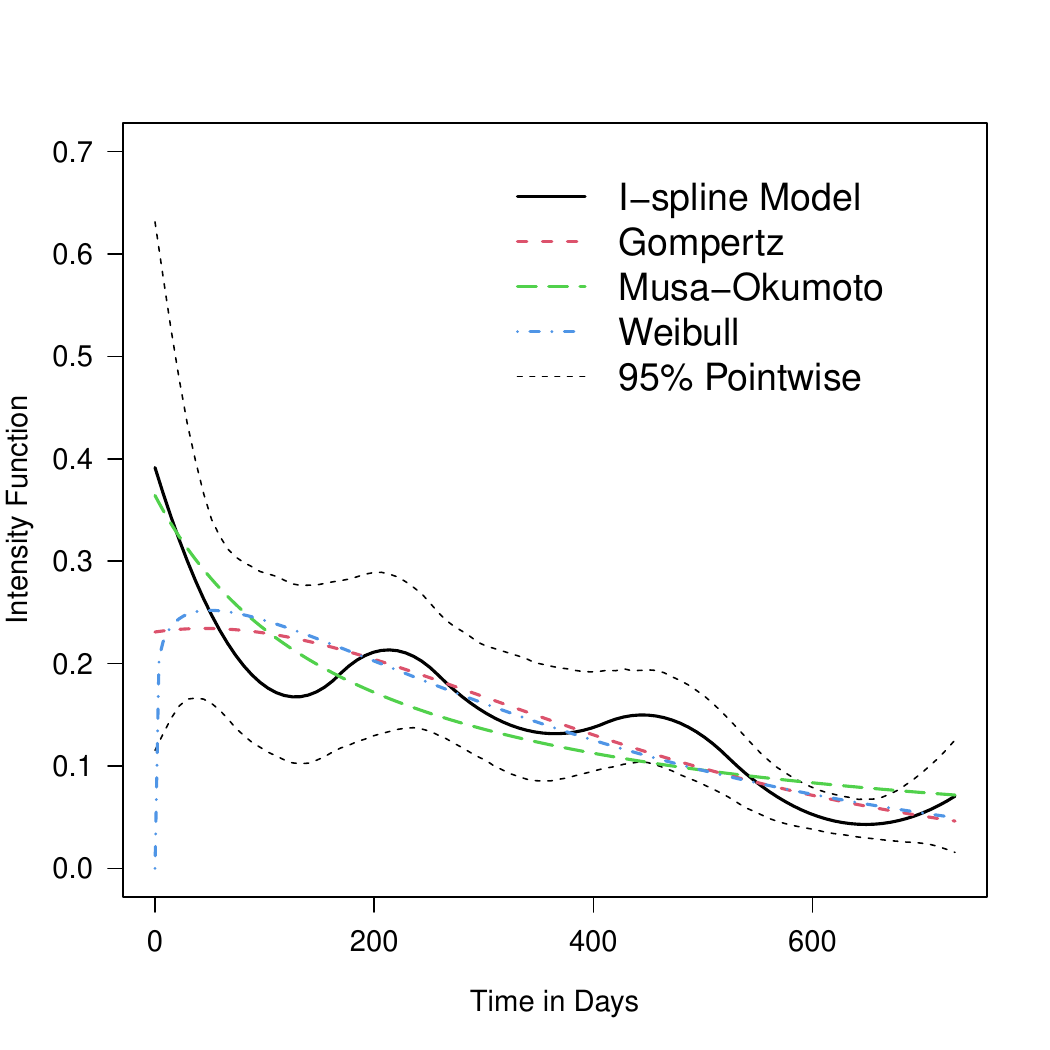}\\
		(a) Waymo & (b) Cruise
	\end{tabular}
	\caption{Estimated BIFs based on parametric models and the I-spline model. \emph{Figure reproduced with permission from Oxford University Press.}}\label{fig:dis.int.est}
\end{figure}

\shortciteN{MinHongKingMeeker2020} employed both parametric and non-parametric methods to model the CBIF and BIF in \eqref{eqn:event.lik}. The Gompertz, Musa-Okumoto, and Weibull models were utilized. Furthermore, a more flexible non-parametric I-spline model was proposed. Figure~\ref{fig:dis.int.est} illustrates the estimated BIF for two manufacturers using both parametric and non-parametric models. Since a decreasing trend in BIF indicates improved AI reliability, the results suggest that AI reliability is improving for Waymo and Cruise. Further details of the analysis can be found in \shortciteN{MinHongKingMeeker2020}.

\subsection{System Level Test Data Set 2}\label{sec:collision}

\subsubsection{Data Description}\label{sec:collision.description}
\begin{figure}
\centering
\includegraphics[width=.95\textwidth]{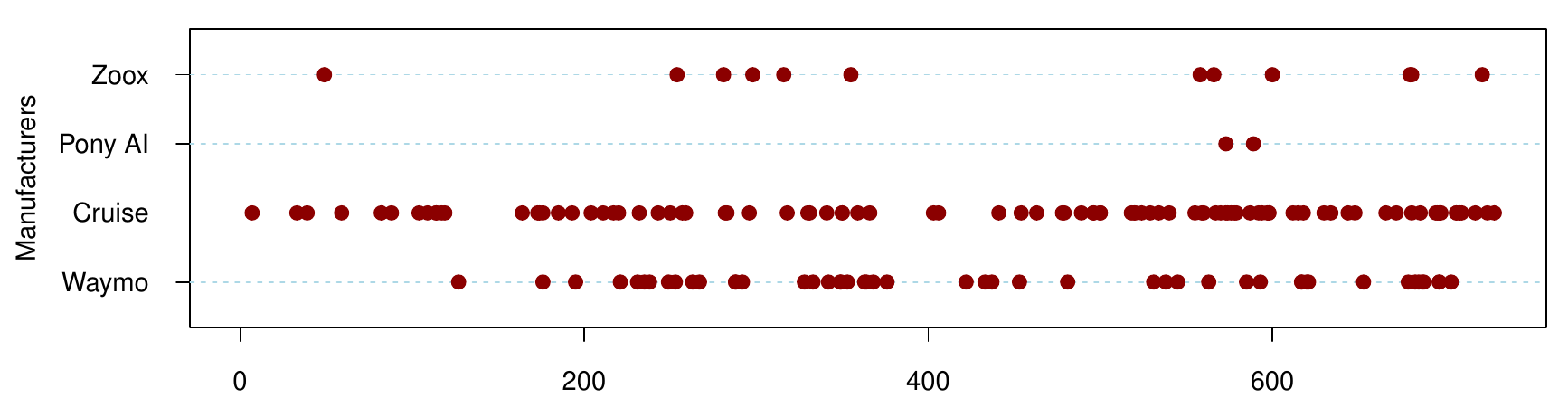}
\caption{Plot of collision events over time (in days) with a dot representing an event.}\label{fig:recurrent.collision}
\end{figure}

In addition to disengagement event data introduced in Section~\ref{sec:disengagement}, another type of recurrent event data that can be used to investigate AV system reliability is collision event data. As the name suggests, collision event data is a type of recurrent event data used to collect information about AV collisions occurring over consecutive time periods for a specific VIN from each manufacturer. Similar to the disengagement events data described in Section~\ref{sec:disengagement}, the collision events data is collected through the AVT program and is published for public review and assessment. The raw collision events data can be downloaded from the \citeN{CAdriving} in PDF format, with separate files available for each manufacturer based on the collision event date. Note that, as of 2024, 11 years of collision events data are publicly available for online download. In terms of data cleaning, it is necessary to extract important information (e.g., manufacturer, collision event date and time, vehicle make, model, and driving mode) into an Excel file for each manufacturer to facilitate further analysis. One important point to mention is that collision event data, as a type of recurrent event data, does not include VIN-level details as described in Section~\ref{sec:disengagement}. Instead, it is available only at the manufacturer level, which is one level higher than the disengagement events data. In addition, we use the same mileage information dataset as described in Section~\ref{sec:disengagement}, which records the monthly mileage information for each AV test unit. Similarly, daily mileage is calculated as the total mileage driven in a month divided by the number of days in that month, as described by \shortciteN{MinHongKingMeeker2020}. We also have a time interval dataset that records the number of days in each month, which can be used for further analysis. A visualization of the available two-year collision event data is shown in Figure~\ref{fig:recurrent.collision}.

\subsubsection{Data Dictionary}
Compared to the disengagement events data described in Section~\ref{sec:disengagement}, more information can be collected and utilized from the collision event data. Additional details about the variables are provided in Table~\ref{tab:collision.var}.

\begin{table}
\caption{Data dictionary for the California DMV collision events dataset.}\label{tab:collision.var}
{\small
\begin{center}
\begin{tabular}{l|l} \hline\hline
Variable & Description\\\hline
Manufacture    & Manufacture of the AVs \\ \hline
VIN           & Unique vehicle identify number                                                                                   \\\hline
Date           & Collision event dates                                                                            \\\hline
Month         & Collision event months                                                                         \\\hline
MonthID         & Identify the numbers for the 24 months in the 2-year period                                                                   \\\hline
EventID       & Identify the number of distinct collision event dates   \\      \hline\hline
\end{tabular}
\end{center}
}
\end{table}

\subsubsection{Data Illustration}
In terms of statistical modeling, a typical application of collision event data is using an NHPP to model recurrent event processes for collision events. Regarding collision events, we can only observe the event times at the manufacturer level; in other words, we do not know which specific vehicle contributed to a given collision event. Let $t_{jk}$ denote the $k$th collision event time for manufacturer $j$, where $k = 1, \ldots, n_{j}$ and $n_{j}$ is the total number of collision events from manufacturer $j$. In addition, let $x_{ij}(t)$ denote the mileage driven by unit $i$ from manufacture $j$, hereafter denoted as unit $(i, j)$, at time $t$ (on a daily basis), where $0<t \leq \tau$ and $\tau=730$ days (i.e., 2 years), representing the duration of the testing period. The event intensity function for unit $(i, j)$ at time $t$ can be modeled as follows:
\begin{align}\label{eqn:event.intensity.func}
\lambda_{ij}(t)=\lambda_{0j}(t;\thetavec)x_{ij}(t),
\end{align}
where $\lambda_{0j}(t;\thetavec)$ denotes the BIF from manufacture $j$ and $\thetavec$ represents the unknown parameters involved in the BIF.

\begin{figure}
	\centering
	\begin{tabular}{cc}
		\includegraphics[width=.4\textwidth]{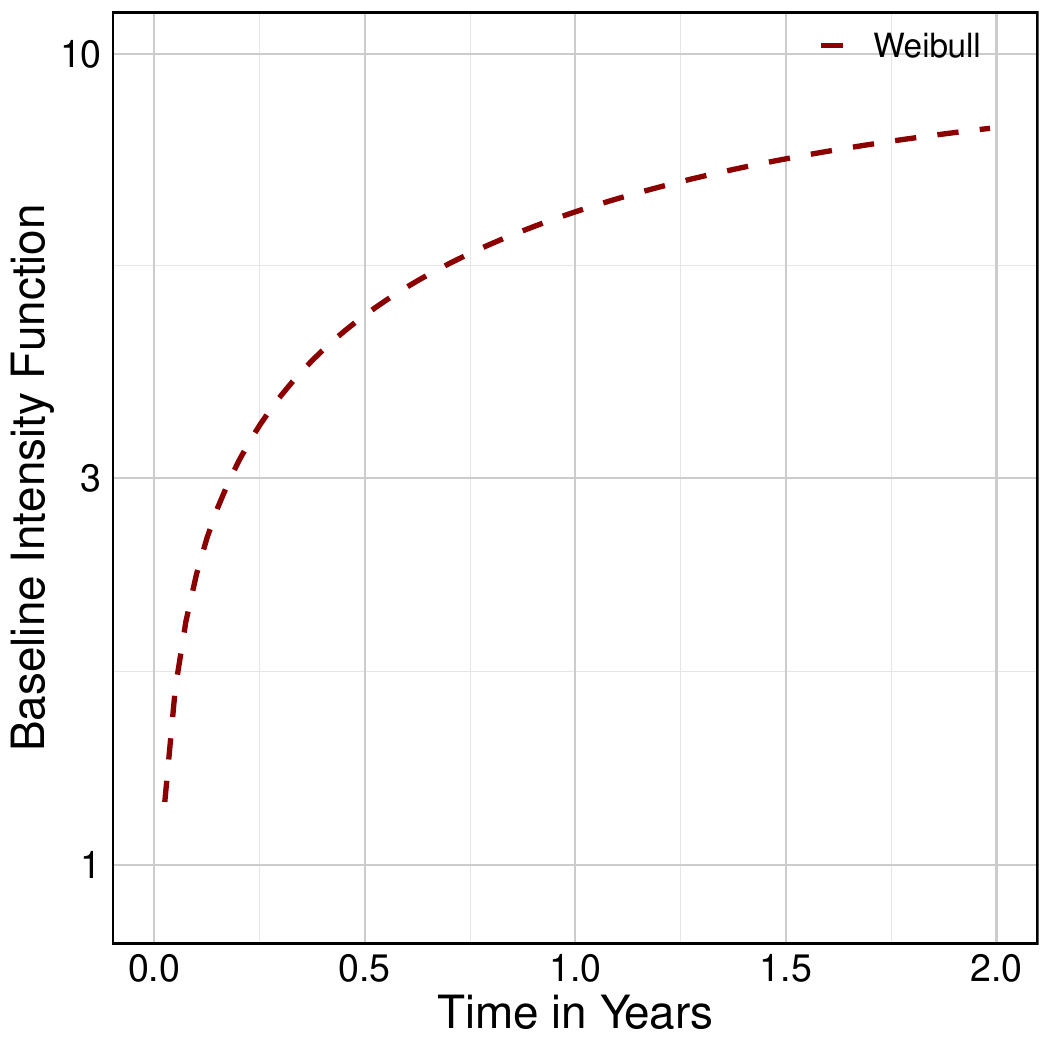}&
		\includegraphics[width=.4\textwidth]{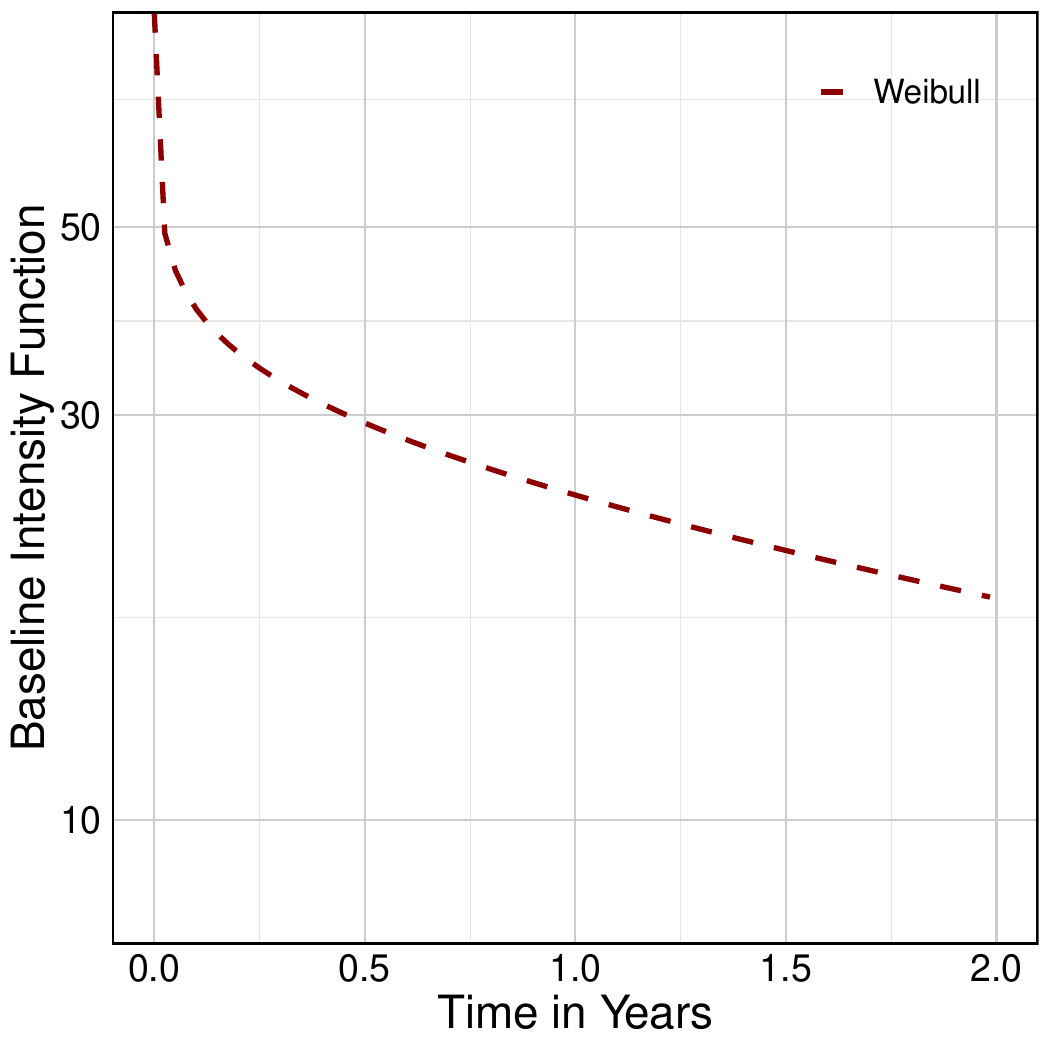}\\
		(a) Waymo & (b) Cruise
	\end{tabular}
	\caption{Estimated BIFs for both manufacturers using the Weibull model.}\label{fig:BIF.collision}
\end{figure}

More specifically, we model the BIF using the Weibull reliability growth model. The specific parametric form of the BIF is as follows:
\begin{align}\label{eqn:Weibull.BIF}
\lambda_{0}(t;\thetavec)=\theta_{1}\theta_{2}\theta_{3}t^{\theta_{3}-1}\exp(-\theta_{2}t^{\theta_{3}}),
\end{align}
where $\theta_{1}>0$, $\theta_{2}>0$, $\theta_{3}>0$, and $\thetavec=(\theta_{1},\theta_{2},\theta_{3})'$.
In addition, the BIF for the Weibull model fitting based on the two-year collision event data is shown in Figure~\ref{fig:BIF.collision}.

\section{The Setup of the DR-AIR Repository}\label{sec:DR.AIR.setup}

The online repository DR-AIR is available at \url{https://github.com/yili-hong/DR-AIR}. It includes a general summary file, \texttt{DataList.csv}, which provides an overview of the datasets in the repository. Each dataset has its own subdirectory. For example, the subdirectory \texttt{AI-Incident-Data-2021} contains files for the AI incident dataset related to reliability, as used in \shortciteN{hong2023statistical}.

Within each dataset's subdirectory, there is a file named \texttt{DataDescription.txt}, which gives the data description. Numerical datasets are stored in .csv format, while other data types, such as images, may be stored as .png files. The data description file provides information on the dataset, including its background, original source, and key details necessary for understanding its variables.

The DR-AIR repository is freely accessible to everyone. The datasets in the DR-AIR repository are subject to the GPL-3.0 license. However, users are encouraged to cite this paper and the original sources of the datasets.

As research on AI reliability progresses, we anticipate adding more datasets to the repository. We encourage the research community to contribute and share AI reliability data to further advance this important field of study. Contact information for the repository maintainer is available online.

\section{Concluding Remarks}\label{sec:concluding.remarks}

This paper focuses on the data aspect of AI reliability research. We discuss key measurements and data types relevant to AI reliability and describe methods for data collection. Emphasizing the importance of applying DoE and ALT principles, we highlight strategies to improve data collection. In addition, we present the datasets gathered for AI reliability research and introduce DR-AIR, an online repository designed to host and share these datasets.

While this work provides valuable insights into the data aspect, several areas warrant further exploration. The modeling of AI reliability can be highly complex, particularly when identifying key predictive factors that influence reliability. Although we reviewed several papers that propose models and analyses for AI reliability, this remains an evolving area with significant challenges and opportunities for future research.

This study has several limitations. AI is an inherently diverse and rapidly advancing field, making it difficult to conduct an exhaustive literature review. Additionally, our focus in this work is primarily on algorithmic performance and some system-level test data. So far, we have not yet seen degradation data in AI reliability, which can also be an important type of reliability data. Furthermore, this paper mainly focuses on the software components of AI systems, leaving hardware considerations largely unaddressed. In modeling of hardware, such as GPU reliability, \shortciteN{Ostrouchovetal2020} and \shortciteN{Min2023-GPU} provide more details on the reliability of hardware components like GPUs.

We conclude this paper with a call to action, urging the research community to contribute to and share AI reliability data. Establishing comprehensive, shared datasets is essential to advancing this critical field, enabling better models, improved methodologies, and a deeper understanding of AI reliability.

\section*{Acknowledgments}

The authors acknowledge the Advanced Research Computing program at Virginia Tech for providing computational resources. The work by Deng and Hong was supported in part by the COS Dean's Discovery Fund at Virginia Tech (Award: 452021). The work by Hong was supported in part by the Data Science Faculty Fellowship (Award: 452118) at Virginia Tech.



\begin{thebibliography}{}

\bibitem[\protect\citeauthoryear{{AI~Incident}}{{AI~Incident}}{2024}]{AIIncidentDB}
{AI~Incident} (2024).
\newblock [{Online}]. {Artificial Intelligence Incident Database:}
  \url{https://incidentdatabase.ai}, accessed: December 26, 2024.

\bibitem[\protect\citeauthoryear{Anderson-Cook and Lu}{Anderson-Cook and
  Lu}{2023}]{anderson2023designed}
Anderson-Cook, C.~M. and L.~Lu (2023).
\newblock Is designed data collection still relevant in the big data era?
\newblock {\em Quality and Reliability Engineering International\/}~{\em
  39\/}(4), 1085--1101.

\bibitem[\protect\citeauthoryear{Antony}{Antony}{2023}]{antony2023design}
Antony, J. (2023).
\newblock {\em Design of experiments for engineers and scientists}.
\newblock Elsevier.

\bibitem[\protect\citeauthoryear{Blood, Herbert, and Wayne}{Blood
  et~al.}{2023}]{blood2023reliability}
Blood, J.~C., N.~W. Herbert, and M.~R. Wayne (2023).
\newblock Reliability assurance for ai systems.
\newblock In {\em 2023 Annual Reliability and Maintainability Symposium
  (RAMS)}, pp.\  1--6. IEEE.

\bibitem[\protect\citeauthoryear{{California DMV}}{{California
  DMV}}{2024}]{CAdriving}
{California DMV} (2024).
\newblock Autonomous vehicle tester program.
\newblock [{Online}]. {Available at:}
  \url{https://www.dmv.ca.gov/portal/vehicle-industry-services/autonomous-vehicles/},
  accessed: September 01, 2024.

\bibitem[\protect\citeauthoryear{Chen, He, Benesty, Khotilovich, and Tang}{Chen
  et~al.}{2015}]{chen2015xgboost}
Chen, T., T.~He, M.~Benesty, V.~Khotilovich, and Y.~Tang (2015).
\newblock Xgboost: extreme gradient boosting.
\newblock {\em R package version 0.4-2\/}, 1--4.

\bibitem[\protect\citeauthoryear{Cody, Lanus, Doyle, and Freeman}{Cody
  et~al.}{2022}]{cody2022systematic}
Cody, T., E.~Lanus, D.~D. Doyle, and L.~Freeman (2022).
\newblock Systematic training and testing for machine learning using
  combinatorial interaction testing.
\newblock In {\em 2022 IEEE International Conference on Software Testing,
  Verification and Validation Workshops (ICSTW)}, pp.\  102--109. IEEE.

\bibitem[\protect\citeauthoryear{Cornell}{Cornell}{2011}]{cornell2011experiments}
Cornell, J.~A. (2011).
\newblock {\em Experiments with mixtures: designs, models, and the analysis of
  mixture data}, Volume 403.
\newblock Hoboken, NJ: John Wiley \& Sons.

\bibitem[\protect\citeauthoryear{{da~Mata}}{{da~Mata}}{2024}]{RRML2025}
{da~Mata}, K. (2024).
\newblock Reliability and resilience of a machine learning model in adversarial
  scenarios.
\newblock \url{https://github.com/karendamata/RRML}.
\newblock Accessed: 2025-01-24.

\bibitem[\protect\citeauthoryear{Dahouda and Joe}{Dahouda and
  Joe}{2021}]{dahouda2021deep}
Dahouda, M.~K. and I.~Joe (2021).
\newblock A deep-learned embedding technique for categorical features encoding.
\newblock {\em IEEE Access\/}~{\em 9}, 114381--114391.

\bibitem[\protect\citeauthoryear{Dosovitskiy, Ros, Codevilla, Lopez, and
  Koltun}{Dosovitskiy et~al.}{2017}]{dosovitskiy2017carla}
Dosovitskiy, A., G.~Ros, F.~Codevilla, A.~Lopez, and V.~Koltun (2017).
\newblock Carla: An open urban driving simulator.
\newblock In {\em Conference on robot learning}, pp.\  1--16. PMLR.

\bibitem[\protect\citeauthoryear{Escobar and Meeker}{Escobar and
  Meeker}{2006}]{escobar2006review}
Escobar, L.~A. and W.~Q. Meeker (2006).
\newblock A review of accelerated test models.
\newblock {\em Statistical science\/}, 552--577.

\bibitem[\protect\citeauthoryear{Faddi, {da~Mata}, Silva, Nagaraju, Ghosh, Kul,
  and Fiondella}{Faddi et~al.}{2024}]{Faddietal2024}
Faddi, Z., K.~{da~Mata}, P.~Silva, V.~Nagaraju, S.~Ghosh, G.~Kul, and
  L.~Fiondella (2024).
\newblock Quantitative assessment of machine learning reliability and
  resilience.
\newblock {\em Risk Analysis, in press\/}.

\bibitem[\protect\citeauthoryear{Fellows and Liu}{Fellows and
  Liu}{2021}]{fellows2021research}
Fellows, R.~F. and A.~M. Liu (2021).
\newblock {\em Research methods for construction}.
\newblock John Wiley \& Sons.

\bibitem[\protect\citeauthoryear{Freeman}{Freeman}{2023}]{freeman2023design}
Freeman, L.~J. (2023).
\newblock Is design data collection still relevant in the big data era? with
  extensions to machine learning.

\bibitem[\protect\citeauthoryear{Goriparthi}{Goriparthi}{2024}]{goriparthi2024ai}
Goriparthi, R.~G. (2024).
\newblock Ai-driven predictive analytics for autonomous systems: A machine
  learning approach.
\newblock {\em Revista de Inteligencia Artificial en Medicina\/}~{\em 15\/}(1),
  843--879.

\bibitem[\protect\citeauthoryear{Gupta and Gupta}{Gupta and
  Gupta}{2022}]{gupta2022research}
Gupta, A. and N.~Gupta (2022).
\newblock {\em Research methodology}.
\newblock SBPD publications.

\bibitem[\protect\citeauthoryear{Hong, Lian, Xu, Min, Wang, Freeman, and
  Deng}{Hong et~al.}{2023}]{hong2023statistical}
Hong, Y., J.~Lian, L.~Xu, J.~Min, Y.~Wang, L.~J. Freeman, and X.~Deng (2023).
\newblock Statistical perspectives on reliability of artificial intelligence
  systems.
\newblock {\em Quality Engineering\/}~{\em 35\/}(1), 56--78.

\bibitem[\protect\citeauthoryear{Hong and Meeker}{Hong and
  Meeker}{2013}]{HongMeeker2013}
Hong, Y. and W.~Q. Meeker (2013).
\newblock Field-failure predictions based on failure-time data with dynamic
  covariate information.
\newblock {\em Technometrics\/}~{\em 55}, 135--149.

\bibitem[\protect\citeauthoryear{Howard, Sirotin, Tipton, and Vemury}{Howard
  et~al.}{2021}]{howard2021reliability}
Howard, J.~J., Y.~B. Sirotin, J.~L. Tipton, and A.~R. Vemury (2021).
\newblock Reliability and validity of image-based and self-reported skin
  phenotype metrics.
\newblock {\em IEEE Transactions on Biometrics, Behavior, and Identity
  Science\/}~{\em 3\/}(4), 550--560.

\bibitem[\protect\citeauthoryear{Inel, Draws, and Aroyo}{Inel
  et~al.}{2023}]{inel2023collect}
Inel, O., T.~Draws, and L.~Aroyo (2023).
\newblock Collect, measure, repeat: Reliability factors for responsible ai data
  collection.
\newblock In {\em Proceedings of the AAAI Conference on Human Computation and
  Crowdsourcing}, Volume~11, pp.\  51--64.

\bibitem[\protect\citeauthoryear{Joseph}{Joseph}{2016}]{joseph2016space}
Joseph, V.~R. (2016).
\newblock Space-filling designs for computer experiments: A review.
\newblock {\em Quality Engineering\/}~{\em 28\/}(1), 28--35.

\bibitem[\protect\citeauthoryear{{Kaggle}}{{Kaggle}}{2025}]{Kaggle}
{Kaggle} (2025).
\newblock Kaggle datasets repository.
\newblock [{Online}]. {Available at:} \url{https://www.kaggle.com/datasets},
  accessed: February 01, 2025.

\bibitem[\protect\citeauthoryear{Karunarathna, Gunasena, Hapuarachchi, and
  Gunathilake}{Karunarathna et~al.}{2024}]{karunarathna2024crucial}
Karunarathna, I., P.~Gunasena, T.~Hapuarachchi, and S.~Gunathilake (2024).
\newblock The crucial role of data collection in research: Techniques,
  challenges, and best practices.
\newblock {\em Uva Clinical Research\/}, 1--24.

\bibitem[\protect\citeauthoryear{Kato, Tokunaga, Maruyama, Maeda, Hirabayashi,
  Kitsukawa, Monrroy, Ando, Fujii, and Azumi}{Kato
  et~al.}{2018}]{kato2018autoware}
Kato, S., S.~Tokunaga, Y.~Maruyama, S.~Maeda, M.~Hirabayashi, Y.~Kitsukawa,
  A.~Monrroy, T.~Ando, Y.~Fujii, and T.~Azumi (2018).
\newblock Autoware on board: Enabling autonomous vehicles with embedded
  systems.
\newblock In {\em 2018 ACM/IEEE 9th International Conference on Cyber-Physical
  Systems (ICCPS)}, pp.\  287--296. IEEE.

\bibitem[\protect\citeauthoryear{Kim}{Kim}{2014}]{kim2014convolutional}
Kim, Y. (2014).
\newblock Convolutional neural networks for sentence classification.
\newblock {\em arXiv:1408.5882\/}.

\bibitem[\protect\citeauthoryear{Koski and Murphy}{Koski and
  Murphy}{2021}]{koski2021ai}
Koski, E. and J.~Murphy (2021).
\newblock Ai in healthcare.
\newblock In {\em Nurses and Midwives in the Digital Age}, pp.\  295--299. IOS
  Press.

\bibitem[\protect\citeauthoryear{Krajzewicz}{Krajzewicz}{2010}]{krajzewicz2010traffic}
Krajzewicz, D. (2010).
\newblock Traffic simulation with sumo--simulation of urban mobility.
\newblock {\em Fundamentals of traffic simulation\/}, 269--293.

\bibitem[\protect\citeauthoryear{Krizhevsky, Hinton, et~al.}{Krizhevsky
  et~al.}{2009}]{krizhevsky2009learning}
Krizhevsky, A., G.~Hinton, et~al. (2009).
\newblock Learning multiple layers of features from tiny images.

\bibitem[\protect\citeauthoryear{Lian, Freeman, Hong, and Deng}{Lian
  et~al.}{2021}]{Lianetal2021Robustness}
Lian, J., L.~Freeman, Y.~Hong, and X.~Deng (2021).
\newblock Robustness with respect to class imbalance in artificial intelligence
  classification algorithms.
\newblock {\em Journal of Quality Technology\/}~{\em 53}, 505--525.

\bibitem[\protect\citeauthoryear{Liu, Xu, Zhang, Muhammad, and Fu}{Liu
  et~al.}{2022}]{liu2022reliable}
Liu, S., X.~Xu, Y.~Zhang, K.~Muhammad, and W.~Fu (2022).
\newblock A reliable sample selection strategy for weakly supervised visual
  tracking.
\newblock {\em IEEE Transactions on Reliability\/}~{\em 72\/}(1), 15--26.

\bibitem[\protect\citeauthoryear{McCullagh and Nelder}{McCullagh and
  Nelder}{1999}]{McCullaghNelder1999}
McCullagh, P. and J.~A. Nelder (1999).
\newblock {\em Generalized Linear Models}.
\newblock FL: Boca Raton: Chapman \& Hall/CRC.

\bibitem[\protect\citeauthoryear{Meeker, Escobar, and Pascual}{Meeker
  et~al.}{2022}]{meeker2022statistical}
Meeker, W.~Q., L.~A. Escobar, and F.~G. Pascual (2022).
\newblock {\em Statistical methods for reliability data}.
\newblock John Wiley \& Sons.

\bibitem[\protect\citeauthoryear{Min, Hong, King, and Meeker}{Min
  et~al.}{2022}]{MinHongKingMeeker2020}
Min, J., Y.~Hong, C.~B. King, and W.~Q. Meeker (2022).
\newblock Reliability analysis of artificial intelligence systems using
  recurrent events data from autonomous vehicles.
\newblock {\em Journal of the Royal Statistical Society: Series C (Applied
  Statistics)\/}~{\em 71}, 987--1013.

\bibitem[\protect\citeauthoryear{Min, Hong, Meeker, and Ostrouchov}{Min
  et~al.}{2023}]{Min2023-GPU}
Min, J., Y.~Hong, W.~Meeker, and G.~Ostrouchov (2023).
\newblock A spatially correlated competing risks time-to-event model for
  supercomputer {GPU} failure data.
\newblock {\em arXiv: 2303.16369\/}.

\bibitem[\protect\citeauthoryear{Mohamed, Khanan, Bashir, Mohamed, Adiel, and
  Elsadig}{Mohamed et~al.}{2024}]{mohamed2024impact}
Mohamed, Y.~A., A.~Khanan, M.~Bashir, A.~H.~H. Mohamed, M.~A. Adiel, and M.~A.
  Elsadig (2024).
\newblock The impact of artificial intelligence on language translation: a
  review.
\newblock {\em Ieee Access\/}~{\em 12}, 25553--25579.

\bibitem[\protect\citeauthoryear{Morris and Mitchell}{Morris and
  Mitchell}{1995}]{morris1995exploratory}
Morris, M.~D. and T.~J. Mitchell (1995).
\newblock Exploratory designs for computational experiments.
\newblock {\em Journal of statistical planning and inference\/}~{\em 43\/}(3),
  381--402.

\bibitem[\protect\citeauthoryear{Nagaraju, Jayasinghe, and Fiondella}{Nagaraju
  et~al.}{2020}]{2020nagaraju}
Nagaraju, V., C.~Jayasinghe, and L.~Fiondella (2020).
\newblock Optimal test activity allocation for covariate software reliability
  and security models.
\newblock {\em Journal of Systems and Software\/}~{\em 168}, 110643.

\bibitem[\protect\citeauthoryear{Nawaz}{Nawaz}{2020}]{nawaz2020artificial}
Nawaz, N. (2020).
\newblock Artificial intelligence applications for face recognition in
  recruitment process.
\newblock {\em Journal of Management Information and Decision Sciences\/}~{\em
  23}, 499--509.

\bibitem[\protect\citeauthoryear{Ostrouchov, Maxwell, Ashraf, Engelmann,
  Shankar, and Rogers}{Ostrouchov et~al.}{2020}]{Ostrouchovetal2020}
Ostrouchov, G., D.~Maxwell, R.~A. Ashraf, C.~Engelmann, M.~Shankar, and J.~H.
  Rogers (2020).
\newblock {GPU} lifetimes on {Titan} supercomputer: Survival analysis and
  reliability.
\newblock In {\em Proceedings of the International Conference for High
  Performance Computing, Networking, Storage and Analysis (SC'20)}, New York,
  NY. Association for Computing Machinery.

\bibitem[\protect\citeauthoryear{Pan, Zhang, Head, Liu, Elli, and Alvarez}{Pan
  et~al.}{2022}]{pan2022quantifying}
Pan, F., Y.~Zhang, L.~Head, J.~Liu, M.~Elli, and I.~Alvarez (2022).
\newblock Quantifying error propagation in multi-stage perception system of
  autonomous vehicles via physics-based simulation.
\newblock In {\em 2022 Winter Simulation Conference (WSC)}, pp.\  2511--2522.
  IEEE.

\bibitem[\protect\citeauthoryear{Pan, Zhang, Liu, Head, Elli, and Alvarez}{Pan
  et~al.}{2024}]{Panetal2024}
Pan, F., Y.~Zhang, J.~Liu, L.~Head, M.~Elli, and I.~Alvarez (2024).
\newblock Reliability modeling for perception systems in autonomous vehicles: A
  recursive event-triggering point process approach.
\newblock {\em Transportation Research Part C: Emerging Technologies\/}~{\em
  169}, 104868.

\bibitem[\protect\citeauthoryear{Shibata, Rinsaka, and Dohi}{Shibata
  et~al.}{2006}]{2006shibata}
Shibata, K., K.~Rinsaka, and T.~Dohi (2006).
\newblock Metrics-based software reliability models using non-homogeneous
  poisson processes.
\newblock In {\em 17th \relax{IEEE} International Symposium on Software
  Reliability Engineering}, pp.\  52--61.

\bibitem[\protect\citeauthoryear{Silva, Hidalgo, Hotchkiss, Dharmasena, Linkov,
  and Fiondella}{Silva et~al.}{2024}]{Silva2024}
Silva, P., M.~Hidalgo, M.~Hotchkiss, L.~Dharmasena, I.~Linkov, and L.~Fiondella
  (2024).
\newblock Predictive resilience modeling using statistical regression methods.
\newblock {\em Mathematics\/}~{\em 12\/}(15).

\bibitem[\protect\citeauthoryear{Smith}{Smith}{2021}]{smith2021reliability}
Smith, D.~J. (2021).
\newblock {\em Reliability, maintainability and risk: practical methods for
  engineers}.
\newblock Butterworth-Heinemann.

\bibitem[\protect\citeauthoryear{Soori, Arezoo, and Dastres}{Soori
  et~al.}{2023}]{soori2023artificial}
Soori, M., B.~Arezoo, and R.~Dastres (2023).
\newblock Artificial intelligence, machine learning and deep learning in
  advanced robotics, a review.
\newblock {\em Cognitive Robotics\/}~{\em 3}, 54--70.

\bibitem[\protect\citeauthoryear{Taufique, Minnehan, and Savakis}{Taufique
  et~al.}{2020}]{taufique2020benchmarking}
Taufique, A. M.~N., B.~Minnehan, and A.~Savakis (2020).
\newblock Benchmarking deep trackers on aerial videos.
\newblock {\em Sensors\/}~{\em 20\/}(2), 547.

\bibitem[\protect\citeauthoryear{Thomas, Martin, Etnier, and Silverman}{Thomas
  et~al.}{2022}]{thomas2022research}
Thomas, J.~R., P.~Martin, J.~L. Etnier, and S.~J. Silverman (2022).
\newblock {\em Research methods in physical activity}.
\newblock Human kinetics.

\bibitem[\protect\citeauthoryear{{UC Irvine}}{{UC Irvine}}{2025}]{UCIRepo}
{UC Irvine} (2025).
\newblock {UC} {Irvine} machine learning repository.
\newblock [{Online}]. {Available at:} \url{https://archive.ics.uci.edu/},
  accessed: February 01, 2025.

\bibitem[\protect\citeauthoryear{Wang, Fu, Du, Gao, Huang, Liu, Chandak, Liu,
  Van~Katwyk, Deac, et~al.}{Wang et~al.}{2023}]{wang2023scientific}
Wang, H., T.~Fu, Y.~Du, W.~Gao, K.~Huang, Z.~Liu, P.~Chandak, S.~Liu,
  P.~Van~Katwyk, A.~Deac, et~al. (2023).
\newblock Scientific discovery in the age of artificial intelligence.
\newblock {\em Nature\/}~{\em 620\/}(7972), 47--60.

\bibitem[\protect\citeauthoryear{Wang, Zhang, Huang, and Zhao}{Wang
  et~al.}{2020}]{wang2020safety}
Wang, J., L.~Zhang, Y.~Huang, and J.~Zhao (2020).
\newblock Safety of autonomous vehicles.
\newblock {\em Journal of advanced transportation\/}~{\em 2020\/}(1), 8867757.

\bibitem[\protect\citeauthoryear{Werner and Schumeg}{Werner and
  Schumeg}{2022}]{werner2022leveraging}
Werner, B. and B.~Schumeg (2022).
\newblock Leveraging traditional design for reliability techniques for
  artificial intelligence.
\newblock In {\em 2022 Annual Reliability and Maintainability Symposium
  (RAMS)}, pp.\  1--6. IEEE.

\bibitem[\protect\citeauthoryear{Zheng, Lu, Hong, and Liu}{Zheng
  et~al.}{2023}]{Zheng2023-testplan}
Zheng, S., L.~Lu, Y.~Hong, and J.~Liu (2023).
\newblock Planning reliability assurance tests for autonomous vehicles.
\newblock {\em arXiv: 2312.00186\/}.

\end{thebibliography}

\end{document}